\documentclass[prd,superscriptaddress,amsfonts,amssymb,onecolumn,amsmath,showpacs]{revtex4-2}
\usepackage{bm}
\usepackage{amsfonts}
\usepackage{latexsym}
\usepackage[utf8]{inputenc}
\usepackage{graphicx}
\usepackage{float}
\usepackage{caption}
\usepackage{subcaption}
\usepackage{adjustbox}
\usepackage{amsmath}
\usepackage{palatino}
\usepackage{mathpazo}
\usepackage[british]{babel}
\usepackage{hhline}
\usepackage{multirow}
\usepackage{textcomp}
\usepackage{float}
\usepackage{booktabs}
\usepackage{dcolumn}
\usepackage{lipsum} 
\usepackage{mdframed}
\usepackage[]{mdframed}
\usepackage{hhline}
\usepackage{multirow}
\usepackage{ragged2e}
\usepackage{hyperref}
\hypersetup{colorlinks,citecolor=blue}
\hypersetup{colorlinks=true,linkcolor=red,filecolor=magenta,    urlcolor=cyan}
\usepackage{amsmath}
\usepackage{xcolor}
\usepackage{orcidlink}
\usepackage{epsfig}
\usepackage{caption}
\usepackage{subcaption}
\usepackage{commath}

\def\jnl@style{\it}
\def\aaref@jnl#1{{\jnl@style#1}}

\def\aaref@jnl#1{{\jnl@style#1}}

\def\aj{\aaref@jnl{AJ}}                   
\def\apj{\aaref@jnl{ApJ}}                 
\def\apjl{\aaref@jnl{ApJ}}                
\def\apjs{\aaref@jnl{ApJS}}               
\def\apss{\aaref@jnl{Ap\&SS}}             
\def\aap{\aaref@jnl{A\&A}}                
\def\aapr{\aaref@jnl{A\&A~Rev.}}          
\def\aaps{\aaref@jnl{A\&AS}}              
\def\mnras{\aaref@jnl{Mon.~Not.~Roy.~Astron.~Soc.}}             
\def\prd{\aaref@jnl{Phys.~Rev.~D}}        
\def\prc{\aaref@jnl{Phys.~Rev.~C}}  
\def\prl{\aaref@jnl{Phys.~Rev.~Lett.}}    
\def\qjras{\aaref@jnl{QJRAS}}             
\def\skytel{\aaref@jnl{S\&T}}             
\def\ssr{\aaref@jnl{Space~Sci.~Rev.}}     
\def\zap{\aaref@jnl{ZAp}}                 
\def\nat{\aaref@jnl{Nature}}              
\def\aplett{\aaref@jnl{Astrophys.~Lett.}} 
\def\apspr{\aaref@jnl{Astrophys.~Space~Phys.~Res.}} 
\def\physrep{\aaref@jnl{Phys.~Rep.}}      
\def\physscr{\aaref@jnl{Phys.~Scr}}       
\def\commat{\aaref@jnl{Comm.~Math.~Phys.}}              
\def\science{\aaref@jnl{Science}}               
\def\cqg{\aaref@jnl{Classical Quant.~Grav.}}            
\def\jpcs{\aaref@jnl{JPCS}}                                     
\def\ijmpd{\aaref@jnl{Int.~J.~Mod.~Phys.~D}}                    
\def\grg{\aaref@jnl{Gen.~Relat.~Gravit.}}               
\def\rpp{\aaref@jnl{Rep.~Prog.~Phys.}}          
\def\npa{\aaref@jnl{Nucl.~Phys.~A}}        
\def\lrr{\aaref@jnl{Living Rev.~Rel.}}                   
\def\jcap{\aaref@jnl{J.~Cosmology Astropart.~Phys.}}    
\def\rmp{\aaref@jnl{Rev.~Mod.~Phys.}}   
\def\epjc{\aaref@jnl{Eur.~Phys.~J.~C}} 
\def\plb{\aaref@jnl{~Phy.~Lett.~B}} 
\def\mpla{\aaref@jnl{Mod.~Phy.~Lett.~A}} 
\def\arxiv{\aaref@jnl{arxiv.org}}

\allowdisplaybreaks[1]

\begin{document}

\title{Teleparallel Gravity and Quintessence: The Role of Nonminimal Boundary Couplings}

\author{S. A. Kadam\orcidlink{0000-0002-2799-7870}}
\email{k.siddheshwar47@gmail.com}
\affiliation{Department of Mathematics, Birla Institute of Technology and Science-Pilani, Hyderabad Campus, Hyderabad-500078, India}

\author{L. K. Duchaniya \orcidlink{0000-0001-6457-2225}}
\email{duchaniya98@gmail.com}
\affiliation{Department of Mathematics, Birla Institute of Technology and Science-Pilani, Hyderabad Campus, Hyderabad-500078, India}

\author{B. Mishra\orcidlink{0000-0001-5527-3565}}
\email{bivu@hyderabad.bits-pilani.ac.in}
\affiliation{Department of Mathematics, Birla Institute of Technology and Science-Pilani, Hyderabad Campus, Hyderabad-500078, India}


\begin{abstract} {\textbf{Abstract:}} 
In this paper, we have outlined the development of an autonomous dynamical system within a general scalar-tensor gravity framework. This framework encompasses the overall structure of the non-minimally coupled scalar field functions for both the torsion scalar ($T$) and the boundary term ($B$). We have examined three well-motivated forms of potential functions and constrained the model parameters through dynamical system analysis. This analysis has played a crucial role in identifying cosmologically viable models. We have analysed the behaviour of dynamical parameters such as equation-of-state parameters, as well all the standard density parameters for radiation, matter, and DE to assess their compatibility with current observational data. The phase space diagrams are presented to support the stability conditions of the corresponding critical points. The Universe is apparent in its late-time cosmic acceleration phase via the DE-dominated critical points. Additionally, we compare our findings with the most prevailing $\Lambda$CDM model. The outcomes are further inspected using the cosmological data sets of Supernovae Ia and the Hubble rate $H(z)$. 

\end{abstract}

\maketitle


\section{Introduction}
Einstein's General Relativity (GR) has successfully explained various astrophysical phenomena, from tests within our solar system to strong field gravitational wave physics \cite{misner1973gravitation}. The observations of galaxies and their dynamical structures suggest the existence of dark matter (DM), which interacts only through gravity and may not fit into the standard model of particle physics \cite{Baudis:2016qwx,Bertone:2004pz}. Moreover, the modification of GR has been significantly influenced by the recent observation of the accelerating expansion of the Universe \cite{Riess:1998cb,Perlmutter:1998np}, which is predicted to be caused by an observational fact known as dark energy (DE). 
The $\Lambda$CDM model, one of the most promising models, proposes that DE is represented by a cosmological constant with a constant Equation of State (EoS) \cite{Sahni_2000}. Although favored by many cosmological observations, this model faces several theoretical \cite{martin2012} and observational difficulties like the recent findings from the Planck collaboration have shown an increasing discrepancy between local and global measurements of $H0$ and $f_{\sigma 8}$ \cite{Aghanim:2018eyx}. It is possible that the issues with $\Lambda$CDM theory may be solved in the upcoming time, or it may be necessary to modify $\Lambda$CDM in some way. Efforts have been made in recent decades to extend GR to address certain aspects of these issues \cite{Capozziello:2011et}. However, it may also be necessary to consider a new approach to address the increasing demands for developing a viable theory of gravity. One such approach is to study the teleparallel gravity theories \cite{bahamonde:2021teleparallel,Cai:2015emx,Clifton:2011jh}.\\

In contrast to GR, the teleparallel equivalent of general relativity (TEGR) offers a different perspective on gravity. TEGR uses a specific connection known as the Weizenb$\ddot{o}$ck connection, which characterizes a space-time with non-zero torsion. TEGR considers torsion a force, while in GR, gravity is understood as the effect of curved space-time. Similar to $f(R)$ extends in GR where $R$ is the Ricci Scalar, TEGR has been further developed by incorporating a function of the torsion scalar $T$, denoted as $f(T)$, into its Lagrangian density \cite{Basilakos:2018arq,Ferraro:2006jd, Cai:2015emx,Bengochea:2008gz}. However, $f(R)$ and $f(T)$ theories lead to different dynamics, as equating $f(R)$ with $f(T)$ and the total derivative is no longer possible. Specifically, $f(R)$ gravity generally yields fourth-order field equations, while $f(T)$ theories are somewhat less problematic as this modification only results in second-order field equations; however, the loss of local Lorentz invariance having concern to this. Researchers in cosmology are studying $f(T)$ gravity \cite{Bamba2012}. The constraints on $f(T)$ cosmology with $Pantheon^{+}$ show potential in explaining aspects of cosmic evolution \cite{MNRASf(T)data}. The spherically symmetric solutions for $f(T)$ gravity models can be obtained using the Noether Symmetry Approach \cite{Paliathanasis_2016}. Furthermore, the validity of $f(T)$ theory  has been supported through the solar system test \cite{Farrugia:2016xcw}.\\

In scalar-tensor theories based on GR, a common coupling function in the corresponding action is represented by a term such as $\xi R \phi^2$ \cite{chernikov1968quantum,PhysRevDBirrell,G.Otalora2013JCAPDSA}. Analogous to this, the same coupling function replacing the $R$ by $T$ has been studied in TEGR \cite{Geng_2012,Xu_2012,WEI2012430,Kucukakca_2013}. These considerations are well suited to describing the behaviour of EoS parameter around the value $-1$ at the present time \cite{Xu_2012}. In TEGR, interesting scenarios are represented in a dynamic model resulting from the non-minimal coupling between the quintessence field similar to the canonical scalar field and $T$ \cite{Geng_2011,MISHRA2024138968,Duchaniya_2023nor}. This action was further modified and analysed considering the non-minimal coupling to the boundary term $B$ in the teleparallel quintessence formalism \cite{Bahamonde_2015}. This formalism can be generalised by replacing the coupling coefficients of $T$ and the boundary term $B$ with a general function of the scalar field $f(\phi), g(\phi)$, in which the generalised second law of thermodynamics has been studied in \cite{Zubair_2017}. Also, the Lorentzian wormholes are constructed using the Noether symmetry \cite{Bahamonde_2016}, which helps to establish some important cosmological
solutions for the modified field equations \cite{Gecim_2018}. The linear stability technique is used to analyse the dynamical properties of the current tachyonic DE model. This analysis involves different potentials and couplings, considering a generalized non-minimal coupling of a tachyonic scalar field with the teleparallel boundary term 
 \cite{Bahamonde_2019}. To study the cosmological dynamics of tachyonic teleparallel DE, one can refer to \cite{Otalora_2013, Banijamali_2012}. \\

 Cosmology in modified gravity theories often leads to complicated systems of differential equations with uncertain initial conditions, making it difficult to perform an analytic treatment. The dynamical system theory is therefore necessary for performing a qualitative analysis. This approach allows us to study the system's long-term behavior and identify regions of instability and attractors. It can also provide us with insight into the dynamics and behaviour of the system. Although the cosmological systems can describe different phases of evolution, at late times, their asymptotic behavior is convergent. It can be determined by a stable critical point obtained from an autonomous system derived from cosmological equations, and the intermediate other eras are described by unstable nodes or saddle points of the same autonomous system. In this work, we focus on constructing and studying an autonomous dynamical system in one of the most general forms of scalar-tensor teleparallel formalism. The comprehensive study of the dynamical system analysis approach in different modified teleparallel gravity formalism has been presented in \cite{Gonzalez2024abc,Rodriguez_Benites_2024, Millano_2024ab,  Hussain_2023, Kadam:2022lgq,KADAM2024Aop,Kadamdynamicalftb,Duchaniya:2022rqu,DUCHANIYA2024101402,Bamba:2012cp,Paliathanasis:2021nqa,Leon:2022oyy,paliathanasis2021epjp}. In this paper, we follow the approach in which we can analyse the dynamics of this scalar tensor formalism in a possible general way. In this formalism, the nonminimal coupling to $T$ and boundary term $B$ is demonstrated and is constructed in \cite{Bahamonde_2015} and in a more general way in \cite{Zubair_2017,Bahamonde_2016,Gecim_2018}. The presented work establishes the dynamical system analysis formalism with the case where the coupling functions of $T$ and $B$ are the general functions of the Canonical scalar field $\phi$. Moreover, in this theory construction, we came across an important point that this formalism obeys the conservation equation. In constructing the dynamical system, we adhere to the exponential coupling function of the scalar field $\phi$ to the boundary term $B$ in the absence of the nonminimal coupling function of $T$. The well-motivated different functional forms of the potential function $V(\phi)$ have been analysed. The paper is organized as follows:  
In Sec. \ref{sec:intro_tg}, we introduce the generalised teleparallel gravity with the non-minimal couplings with $T$, $B$ and derive the corresponding evolution equations.  
In Sec. \ref{exponentialcoupling}, we conduct a dynamical system analysis of the DE model and reveal the fundamental discussion of the critical points.  
Moving on to Sec. \ref{P_1} to Sec. \ref{P_3}, we employ analysis of different scalar field potentials. The numerical approach is used to study the evolution of the field equations for the model and analyse the implications of $T$ and $B$ couplings for the study of the DE EoS parameters. We examined the evolution of the Hubble rate with respect to redshift ($z$) and found that it aligns with the Hubble rate ($H_{\Lambda CDM}(z)$) in the $\Lambda$CDM model. This alignment was validated using Hubble data points \cite{Moresco_2022_25}. We also considered the modulus function $\mu_{\Lambda CDM}$ of the $\Lambda$CDM model, 1048 pantheon data points, and the modulus function $\mu(z)$. Lastly, in Sec. \ref{Conclusion}, we summarize the results and conclusions with the final remarks.

\section{Teleparallel Gravity Formalism}\label{sec:intro_tg}

In this section, we have presented the basic equations and formalism to construct the TEGR, as well as its scalar torsion formalism. In this theory, the tetrad $e^{a}_{~~\mu}$ and its inverses $E_{a}^{~~\mu}$ serve as the dynamical variable \cite{Krssak:2018ywd,Cai:2015emx,Aldrovandi:2013wha}. The Greek indices represent spacetime indices, and the Latin indices denote the tangent space indices. The metric $g_{\mu \nu}$ is expressed in terms of the tetrad as,
\begin{align}\label{metric_tetrad_rel}
    e^{a}_{~~\mu} e^{b}_{~~\nu}\eta_{ab}=g_{\mu\nu}\,,& &  E_{a}^{\ \ \mu} E_{b}^{\ \ \nu}g_{\mu\nu}=\eta_{ab} \,.
\end{align}
Where $\eta_{a b}=\text{diag} \left(-1, 1, 1, 1\right)$ represents the Minkowski space-time metric. Like to the metric, the tetrads meet orthogonality conditions that are in the form of
\begin{align}
    e^{a}_{\ \ \mu} E_{b}^{\ \ \mu}=\delta^a_b\,,&  & e^{a}_{\ \ \mu} E_{a}^{\ \ \nu}=\delta^{\nu}_{\mu}\,.
\end{align}

Similar to the Levi-Cività connection used in GR, the Weitzenb\"{o}ck connection ($\Gamma^{\alpha}_{\, \, \, \nu\mu} $) is chosen in TG \cite{Weitzenbock1923, Krssak:2015oua}. The Weitzenb\"{o}ck  connection is the  teleparallel connection in the Weitzenb\"{o}ck gauge, where the spin connection vanishes \cite{bahamonde:2021teleparallel}. The Weitzenb\"{o}ck connection is central to teleparallel gravity, where gravity is described by torsion rather than curvature, and thus all the information of the gravitational field is
embedded in the torsion tensor, while the gravitational Lagrangian is the torsion scalar. In this framework, gravity is interpreted as a force resulting from torsion in spacetime, unlike the standard curvature interpretation in GR. The Weitzenb\"{o}ck connection allows for a global frame of reference in which the connection coefficients vanish, leading to a teleparallel structure on the manifold. This property allows all vectors to be transported parallel to themselves, leading to a formulation of gravity without curvature. The Weitzenb\"{o}ck connection is expressed in terms of the derivatives of the tetrad fields, which relate the coordinate basis to a locally flat Minkowski space basis.
\begin{equation}
   \Gamma^{\alpha}_{\, \, \, \nu\mu} := E_{a}^{~~\alpha}\left(\partial_{\mu}e^{a}_{~~\nu} +  \omega^{a}_{\, \, \, b\mu} e^{b}_{~~\nu}\right)\,,
\end{equation}
  where $\omega^{a}_{\,\,\, b \mu}$ is a flat spin connection in the theory that incorporates Lorentz transformation invariance, which arises explicitly from the tangent space indices, contrasted with GR, whose spin connections are not flat \cite{misner1973gravitation} due to their tetrads. Tetrad-spin connections in TG represent gravitational and local degrees of freedom, and both determine the equation of motion of the system. To maintain the covariance of the field equations, the spin connection is explicitly included in the Weitzenb\"{o}ck connection \cite{Krssak:2015oua,Hohmann_2018_97_covariant}. 
The torsion tensor $T^{\alpha}_{\ \ \mu \nu}$ in teleparallel gravity is defined via the Weitzenb\"{o}ck connection:
\begin{equation}\label{eq:torsion_tensor}
T^{\alpha}_{\ \ \mu \nu}\equiv\Gamma^{\alpha}_{~~\nu \mu}-\Gamma^{\alpha}_{~~\mu \nu}=E^{~~\alpha}_{a} \partial_{\mu} e^{a}_{~~\nu}-E^{~~\alpha}_{a} \partial_{\nu} e^{a}_{~~\mu}.
\end{equation}
The superpotential $S_{\rho}^{~~\mu \nu}$ can be define as,
\begin{equation}\label{eq:superpotential}
S_{\rho}^{~~\mu \nu}\equiv\frac{1}{2}(K^{\mu \nu}_{~~~\rho}+\delta^{\mu}_{\rho}T^{\alpha \nu}_{~~~\alpha}-\delta^{\nu}_{\rho}T^{\alpha \mu}_{~~~\alpha})\,.    
\end{equation}
Where $K^{\mu \nu}_{~~~\rho}\equiv \frac{1}{2}(T^{\nu \mu}_{~~~\rho}+T_{\rho}^{~~\mu \nu}-T^{\mu \nu}_{~~~\rho})$ represent the contortion tensor. The torsion scalar can be obtained by using the contractions of the torsion tensor as follows,
\begin{equation}\label{eq:torsion_scalar_def}
   T \equiv \frac{1}{4}T^{\alpha}_{\ \ \mu\nu}T_{\alpha}^{\ \ \mu\nu} + \frac{1}{2}T^{\alpha}_{\ \ \mu\nu}T^{\nu\mu}_{\ \ \  \  \alpha} - T^{\beta}_{\ \ \beta\mu}T_{\nu}^{\ \ \nu\mu}\,.
\end{equation}

In the context of exchanging the Levi-Civita connection with the teleparallel connection, it is observed that measures of curvature identically vanish, such as $R\equiv 0$. In teleparallel theories, $T$ can be considered a counterpart to the curvature $R$ in GR. For more details about teleparallel theory, please see reference \cite{Aldrovandi:2013wha,bahamonde:2021teleparallel}. In this scenario, we can express the following relation for the curvature and torsion scalars \cite{Bahamonde:2015zma, Farrugia:2016qqe}.

\begin{equation}\label{LC_TG_conn}
     -T + {2}e^{-1}\partial_{\mu}(eT^{\alpha\mu}_{~~~\alpha})=-T + B=R\,.
\end{equation}
 This relation includes $R$, $T$, and $B ={2}e^{-1}\partial_{\mu}(eT^{\alpha\mu}_{~~~\alpha})$ represents a total divergence term, and $e=\det\left(e^{a}_{~~\mu}\right)=\sqrt{-g}$ stands for the determinant of the tetrad. This ensures that the GR and TEGR actions will produce identical field equations.\\

Initially, to study the late time cosmic acceleration, the works are done in the quintessence scalar field where the power law coupling potentials are considered \cite{Wetterich1988,Ratra1988}, for review in curvature formalism one can refer to \cite{Copeland:2006wr}. The coupling coefficient $\xi \phi^2$ of $T$ was considered in the teleparallel modifications and was analysed in \cite{WEI2012430,Geng:2011}. To analyse the scaling attractors in the teleparallel gravity formalism, the $\xi \phi^2$ coupling coefficient is generalized to $\xi f(\phi)$ in \cite{G.Otalora2013JCAPDSA}. Moreover, in \cite{Bahamonde_2015}, this formalism has further modified the teleparallel approach by considering the inclusion of the nonminimal coupling $\xi \phi^2$, $\chi \phi^2$ to the $T$, and $B$ respectively. The more general form where the general scalar field functions $f(\phi), g(\phi)$ are incorporated to study the generalised second law of thermodynamics and the Noether symmetry approach in \cite{Zubair_2017,Gecim_2018}. Moreover, the generalized non-minimal coupling of a tachyonic
 scalar field with the teleparallel boundary term is studied in \cite{Bahamonde_2019}. The  Lorentzian wormholes are constructed in this formalism by Noether symmetries \cite{Bahamonde_2016}. During the literature study, we came across the scope of the study of dynamical system analysis, which has not been studied previously in this general scalar-tensor formalism. Hence in this study, we aim to construct an autonomous dynamical system to analyse the viability of the different well-motivated scalar field potentials to analyse the different evolution epochs of the Universe. We consider the action formula in which the scalar field is non-minimally coupled to both $T$ and $B$ as follows,

\begin{equation}\label{Scalar_Torsion_Lagrangian}
    \mathcal{S} =\int \left[ \frac{-T}{2 \kappa^2} - \frac{1}{2}\left(f(\phi) T+g(\phi)B - \partial_{\mu} \phi \partial^{\mu} \phi\right) -V(\phi)+ \mathcal{S}_{\text{m}}+\mathcal{S}_{\text{r}}\right]e\, \mathrm{d}^4 x \,.\\
\end{equation}
In modified teleparallel gravity, the gravitational field equations and their solutions rely on the spin connection. Therefore, it is essential to establish a procedure for determining the specific spin connection corresponding to each tetrad field to effectively solve the field equations. In the realm of Friedmann-Lemaître-Robertson-Walker(FLRW) cosmologies, it has been shown that the diagonal tetrad serves as a suitable representation and is expressed as follows,
\begin{equation}\label{flrw_tetrad}
    e^{a}_{\ \ \mu}=\textrm{diag}(1,a(t),a(t),a(t))\,,
\end{equation}
The appropriate spin connection associated with this tetrad is the vanishing spin connection, leading to physically meaningful outcomes\cite{Krssak:2015oua, Hohmann_2018_97_covariant, Gonzalez_Espinoza_2020}. This tetrad choice leads to the flat FLRW metric as,
\begin{equation}
    ds^2 = -dt^2+a(t)^2(dx^2+dy^2+dz^2)\,.
\end{equation}
The tetrad field allows for expressing the $T$ and $B$ in terms of the scale factor and its time derivatives as,
\begin{equation}\label{T, B}
    T =6H^2\,,\quad B=6(3H^2 + \dot{H})\,,
\end{equation}
 where $H \equiv \frac{\dot{a}}{a}$ is the Hubble parameter, a dot represents the derivative
with respect to time. To obtain the contribution of the DE for pressure and the energy density, we use the general Friedmann equations which can be written as,
\begin{align}
3H^2=\kappa^2\left(\rho_{m}+\rho_{r}+\rho_{DE}\right) \label{FridE1}\,,\\
3H^2+2\dot{H}=-\kappa^2\left(p_{r}+p_{DE}\right)\label{FridE2}.
\end{align}

Varying the above action Eq. \eqref{Scalar_Torsion_Lagrangian}, the field equations can be obtained as,
\begin{align}
   3H^2 =& \kappa^2\left[\rho_{m}+\rho_{r}-3H^2 f(\phi)+3Hg^{'}(\phi)\dot{\phi}+V(\phi)+\frac{\dot{\phi}^2}{2}\right], \label{1stFE}\\ 
   3H^2 + 2\dot{H} =&-\kappa^2 \left[ p_{r} -V(\phi)+\left(3H^2 +2\dot{H}\right)f(\phi) +2H f^{'}(\phi) \dot{\phi} +\frac{\dot{\phi}^2}{2}-\dot{\phi}^2 g^{''}(\phi) -g^{'}(\phi) \ddot{\phi}\right] \label{2ndFE}\,.
\end{align}

Where prime denotes the differentiation with respect to the scalar field $\phi$. With the same setting, the Klein-Gordon equation can be obtained as,

\begin{align}
\ddot{\phi}+3H\dot{\phi}+\left(\frac{B}{2} g^{'} (\phi) +\frac{T}{2} f^{'} (\phi) \right)+V^{'} (\phi)=0\,.
\end{align}

Comparing Eqs. \eqref{1stFE}-\eqref{2ndFE} and Eqs. \eqref{FridE1}-\eqref{FridE2}, one can retrace the evolution equations required to analyse the dynamics of the DE as,

\begin{align}
    -3H^{2} f(\phi)+3Hg^{'}(\phi) \dot{\phi} +V(\phi) +\frac{\dot{\phi}^2}{2} &=\kappa^2 \rho_{DE}\,,\label{fede1}\\
   -V(\phi)+ \left(3H^2+2\dot{H}\right) f(\phi) +2H f^{'} (\phi) \dot{\phi} +\frac{\dot{\phi}^2}{2} -\dot{\phi}^2 g^{''}(\phi) -g^{'} (\phi) \ddot{\phi} &=\kappa^2 p_{DE}\,.\label{fede2}
\end{align}
One can easily verify that the equations Eqs. \eqref{fede1}-\eqref{fede2} are satisfying the standard conservation equation as stated below,
\begin{align}
\dot{\rho}_{DE}+3H\left(\rho_{DE}+p_{DE}\right)=0\,,
\end{align}
this equation agrees with the energy conservation law and the fluid evolution equations,
\begin{align}
\dot{\rho}_{m}+3H\rho_{m}&=0\,,\nonumber\\
\dot{\rho}_r+4H \rho_{r}&=0\,.\nonumber\\
\end{align}
The standard density parameters for matter, radiation, and the DE can be written as,
\begin{align}
\Omega_{m}=\frac{\kappa^2 \rho_{m}}{3H^2}, \quad \Omega_{r}=\frac{\kappa^2 \rho_{r}}{3H^2}, \quad \Omega_{DE}=\frac{\kappa^2 \rho_{DE}}{3H^2},
\end{align}

which satisfies the constrained equation,
\begin{align}
\Omega_{m}+\Omega_{r}+\Omega_{DE}=1.
\end{align}

\section{ANALYSIS IN NONMINIMAL COUPLING PURELY TO THE
 BOUNDARY TERM}\label{exponentialcoupling}
 We will analyse the scalar field model non-minimally coupled to the boundary term for the case $f(\phi)=0$ and 
 $g(\phi)\ne 0$. For the case  $f(\phi)\ne 0$ and $g(\phi)= 0$, one may see the Ref.\cite{G.Otalora2013JCAPDSA}, where as for the particular case $f(\phi)=\xi$, the detailed analysis is in Ref.\cite{WEI2012430,Geng:2011}. We define the dimensionless variables as follows,
\begin{align}
x=\frac{\kappa \dot{\phi}}{\sqrt{6}H}\,, \quad y=\frac{\kappa \sqrt{V}}{\sqrt{3}H}\,,\quad u=\kappa g^{'}(\phi)\,, \quad \rho=\frac{\kappa \sqrt{\rho_{r}}}{\sqrt{3}H}\,, \quad \lambda=\frac{-V^{'}(\phi)}{ V(\phi)}\,, \quad \Gamma=\frac{V^{''}(\phi)V(\phi)}{V^{'}(\phi)^{2}}\,.\label{generaldynamical variables}
\end{align}

 The first Friedmann Eq. \eqref{FridE1} can be obtained in the form of dynamical variables as,
\begin{align}
1=\Omega_{m}+\rho^{2}+\sqrt{6} x u +y^2 +x^2,
\end{align}
where,
\begin{align}
    \Omega_{\phi}=\sqrt{6} x u +y^2 +x^2.
\end{align}
 To frame an autonomous dynamical system, we have to consider the exponential coupling function to the boundary term. In this case, so we consider the coupling function as, $g(\phi)=g_{0} e^{-\alpha \phi \kappa}$ \cite{Zlatev_1999,Bahamonde_2019}. Now, The dynamical system can be obtained by taking the differentiation of the dimensionless variables with  $N=log(a)$ as,

\begin{align}
\frac{dx}{dN}&=-3 x-\frac{9}{\sqrt{6}} u+ \frac{3 \lambda y^2}{\sqrt{6}}-\left(x +\frac{3 u}{\sqrt{6}}\right)\frac{\dot{H}}{H^2} \,,\nonumber\\
\frac{dy}{dN}&=-\sqrt{\frac{3}{2}} x \lambda y -y \frac{\dot{H}}{H^2},\nonumber\\
\frac{du}{dN}&=-\sqrt{6} \alpha  u x\,,\nonumber\\
\frac{d\rho}{dN}&=-2 \rho -\rho \frac{\dot{H}}{H^2}\,,\nonumber\\
\frac{d\lambda}{dN}&=-\sqrt{6} x \lambda ^2 \left(\Gamma -1\right)\,,\nonumber\\
\label{general_dynamicalsystemcase_I}
\end{align}
where
\begin{align}
    \frac{\dot{H}}{H^2}&=-\frac{\rho ^2+9 u^2+6 \alpha  u x^2+3 \sqrt{6} u x-3 y^2 (\lambda  u+1)+3 x^2+3}{3 u^2+2}\,.\label{HdotbyH^2}
\end{align}
Using Eq. \eqref{HdotbyH^2} into Eq. \eqref{general_dynamicalsystemcase_I}, the autonomous dynamical system can be generated as follows,  
\begin{align}
\frac{dx}{dN}&=\frac{2 x \left(\rho ^2+9 u^2-3 \lambda  u y^2-3 y^2-3\right)+6 x^3 (2 \alpha  u+1)+3 \sqrt{6} u x^2 (2 \alpha  u+3)+\sqrt{6} \left(u \left(\rho ^2-3 y^2-3\right)+2 \lambda  y^2\right)}{6 u^2+4}\,,\nonumber\\
\frac{dy}{dN}&=-\sqrt{\frac{3}{2}} \lambda  x y+\frac{y \left(\rho ^2+9 u^2+6 \alpha  u x^2+3 \sqrt{6} u x-3 y^2 (\lambda  u+1)+3 x^2+3\right)}{3 u^2+2}\,,\nonumber\\
\frac{du}{dN}&=-\sqrt{6} \alpha  u x\,,\nonumber\\
\frac{d\rho}{dN}&=\frac{\rho  \left(\rho ^2+3 \left(u^2+u x \left(2 \alpha  x+\sqrt{6}\right)-\lambda  u y^2+x^2\right)-3 y^2-1\right)}{3 u^2+2}\,,\nonumber\\
\frac{d\lambda}{dN}&=-\sqrt{6} x f\,,
\label{dynamicalsystem}
\end{align}
where $f=\lambda^2 (\Gamma-1)$ and can be written in a generalise way as, $f=\alpha_{1}\lambda^2+\alpha_{2}\lambda+\alpha_{3}$ \cite{Roy_2018}. The potentials in Table \ref{Potentialfunctions} will be used for further study.
\begin{table}[H]
    \centering 
    \begin{tabular}{|c |c |c |c| c| c|} 
    \hline
 \multicolumn{6}{|c|}{\textbf{List of Potential Functions}} \\
    \hline 
    \parbox[c][0.9cm]{0.9cm}{\textbf{Name / Refs.}
    }&\textbf{Potential function} $V(\phi)$ & \textbf{$f$} &  \textbf{$\alpha_{1}$}& $\alpha_{2}$& $\alpha_{3}$\\ [0.5ex] 
    \hline 
    \parbox[c][0.9cm]{0.9cm}{$P_{1}$ \cite{Zlatev_1999}} & $V_{0}e^{-\kappa \phi}$ & $0$&  $0$& $0$ & 
    $0$  \\
    \hline
   \parbox[c][0.9cm]{0.9cm}{$P_{2}$ \cite{Sahni_2000}} &  $Cosh(\xi \phi)-1$ & $-\frac{\lambda^2}{2}+\frac{\xi^2}{2}$ &  $-\frac{1}{2}$& $0$ & $\frac{\xi^2}{2}$ \\
   \hline
   \parbox[c][0.9cm]{0.9cm}{$P_{3}$ \cite{SAHNI2000}} &   $V_{0} Sinh^{-\eta}(\beta \phi)$ & $\frac{-\lambda^2}{\eta}-\eta \beta^2$ &  $\frac{1}{\eta}$& $0$ & $-\eta \beta^2$ \\
   \hline
    \end{tabular}
    \caption{List of Potentials functions with Corresponding $f$ and values of $\alpha_{1},\alpha_{2}, \alpha_{3}.$}
    \label{Potentialfunctions}
\end{table}
In Table \ref{Potentialfunctions}, $V_{0}, \xi, \eta, \beta$ are the constants. We have demonstrated the dynamical analysis for each of the above cases in detail in the following sections.

\subsection{Potential \texorpdfstring{$P_1 \,:  V_{0}e^{-\kappa \phi}$}{}}\label{P_1}
The exponential potential is considered in the literature to study the teleparallel DE scalar field models \cite{Roy_2018,
Gonzalez-Espinoza:2020jss,duchaniya2023dynamical}. This potential also can be considered to study the generalized teleparallel non-minimally coupled tachyonic models in the presence of the boundary term $B$ \cite{Bahamonde_2019}. The critical points throughout this study are the points at which the autonomous dynamical system vanishes and can be obtained through $\frac{dx}{dN}=0,\,\frac{dy}{dN}=0,\,\frac{du}{dN}=0,\,\frac{d\rho}{dN}=0.$ In this case, the value of $f=0$; hence, the above autonomous dynamical system reduces to the four dimensions. The critical points, along with the value of $q, \omega_{tot}$ and the standard density parameters $\Omega_{r}, \, \Omega_{m},\, \Omega_{DE}$ are presented in the following Table \ref{P1criticalpoints}.
\begin{table}[H]
    \centering 
    \begin{tabular}{|c |c |c |c| c| c| c|} 
    \hline\hline 
    \parbox[c][0.9cm]{1.3cm}{\textbf{C. P.}
    }& $ \{ x, \, y, \, u, \, \rho \} $ & $q$ &  \textbf{$\omega_{tot}$}& $\Omega_{r}$& $\Omega_{m}$& $\Omega_{DE}$\\ [0.5ex] 
    \hline\hline 
    \parbox[c][1.3cm]{1.3cm}{$A_{R}$ } &$\{ 0, 0, 0,  1 \}$ & $1$ &  $\frac{1}{3}$& $1$& $0$ & $0$ \\
    \hline
    \parbox[c][1.3cm]{1.3cm}{$B_{R}$ } & $\left\{\frac{2 \sqrt{\frac{2}{3}}}{\lambda }, \, \, \frac{2}{\sqrt{3} \lambda }, \, \, 0, \, \, \sqrt{1-\frac{4}{\lambda^2}}\right\}$ & $1$&  $\frac{1}{3}$& $1-\frac{4}{\lambda ^2}$ & $0$ & $\frac{4}{\lambda ^2}$ \\
    \hline
   \parbox[c][1.3cm]{1.3cm}{$C_{M}$ } &  $\{0,\, \, 0,\, \, 0, \, \, 0\}$ & $\frac{1}{2}$ &  $0$& $0$ & $1$ & $0$\\
   \hline
   \parbox[c][1.3cm]{1.3cm}{$D_{M}$} &   $\left\{\frac{\sqrt{\frac{3}{2}}}{\lambda }, \, \, \frac{\sqrt{\frac{3}{2}}}{\lambda },\, \, 0, \, \, 0\right\}$ & $\frac{1}{2}$ &  $0$& $0$ & $1-\frac{3}{\lambda ^2}$ & $\frac{3}{\lambda ^2}$\\
   \hline
 \parbox[c][1.3cm]{1.3cm}{$E_{DE}$} &  $\left\{\frac{\lambda }{\sqrt{6}}, \, \, \sqrt{1-\frac{\lambda^2}{6}},\, \, 0,\, \, 0\right\}$ & $-1+\frac{\lambda ^2}{2}$ & $-1+\frac{\lambda ^2}{3}$ & $0$ & $0$ & $1$\\
 \hline
 \parbox[c][1.3cm]{1.3cm}{$F_{DE}$} &  $\left\{ 0,\, \, 1, \, \, \frac{\lambda }{3}, \, \, 0\right\}$ & $-1$ &  $-1$ & $0$ & $0$ & $1$\\
 \hline
 \parbox[c][1.3cm]{1.3cm}{$G_{DE}$} &  $\left\{ 0,\, \, 1, \, \, 0, \, \, 0\right\}$ & $-1$ &  $-1$ & $0$ & $0$ & $1$\\
 \hline
 \parbox[c][1.3cm]{1.3cm}{$H_{D}$} &  $\left\{ 1,\, \,0, \, \, 0, \, \, 0\right\}$ & $2$ &  $1$ & $0$ & $0$ & $1$\\
 \hline
    \end{tabular}
    \caption{Critical points analysis: values of $q$,\, $\omega_{tot}$,\, $\Omega_{r}$,\, $\Omega_{m}$ 
 and $\Omega_{DE}$ for potential \ref{P_1} $(P_1)$.}
    \label{P1criticalpoints}
\end{table}

The detailed analysis of the critical points is presented as follows,
\begin{itemize}
\item{\textbf{Radiation-Dominated Critical Points} $A_R, B_R$}: \\
The critical point $A_R$ represents a standard radiation-dominated era with $\Omega_r=1$. The critical point $B_R$ represents a non-standard radiation-dominated era with $\Omega_r=1-\frac{4}{\lambda^2}$. This critical point is a scaling solution with $\Omega_{DE}=\frac{4}{\lambda^2}$. The physical condition $0<\Omega_{DE}<1$, imposes the condition on $|\lambda|>2$. The value of the deceleration parameter $q=1$, the total EoS parameter $\omega_{tot}=\frac{1}{3}$ at both of these critical points; hence, these critical points describe the early time radiation-dominated era of the Universe.\\

\item{\textbf{Matter-Dominated Critical Points} $C_M, D_M$}: \\
The critical points $C_M$ and $D_M$ both are the critical points describing the matter-dominated epochs of the evolution of the Universe. The value of the parameters $q=\frac{1}{2}, \, \omega_{tot}=0$ at both of these critical points and hence describing the early matter-dominated epoch. The critical point $C_M$ is the standard matter-dominated epoch with $\Omega_{m}=1$. At $D_M$, the small amount of standard density parameter for $DE$ $ \Omega_{DE}=\frac{3}{\lambda^2}$ contributes. This critical point is a non-standard matter-dominated critical point with $\Omega_{m}=1-\frac{3}{\lambda^2}$. Moreover, critical point $D_M$ is the scaling solution, the physical condition on $\Omega_{DE}$ obtains the condition on $\lambda$ as, $|\lambda| >\sqrt{3}$. This critical point also appears in the study of the standard quintessence model \cite{copelandLiddle} and the study of the teleparallel DE model \cite{Xu_2012}.

\item{\textbf{DE-Dominated Critical Points} $E_{DE}, F_{DE}, G_{DE}$}: \\
The critical point $E_{DE}$ describes a DE-dominated epoch of the Universe evolution with $\Omega_{DE}=1$. The value of $\omega_{tot}=-1+\frac{\lambda^2}{3}$, which falls in the quintessence regime. This critical point is a late-time attractor solution, and the values of the $q, \omega_{tot}$ show compatibility with the current observation studies. This critical point also exists in the study of standard quintessence \cite{copelandLiddle} and in the teleparallel DE models \cite{Xu_2012}. The critical point $F_{DE}$ corresponds to an accelerating Universe with complete DE domination ($\omega_{tot}=-1$). The DE behaves like a cosmological constant. This critical point is a novel critical point that is not present in standard quintessence \cite{copelandLiddle}, and also, in terms of the coordinates, it varies from \cite{Xu_2012}. Similar to $F_{DE}$, the critical point $G_{DE}$ is also behaving like a cosmological constant. This solution describes a standard DE-dominated era with $\Omega_{DE}=1$.\\

 \item{\textbf{Critical Point Representing the Stiff DE $H_D$ :}}
  In the stiff matter era, $\omega=\frac{p}{\rho}=1$ and the energy density \(\rho\) evolves as \(\rho \propto a(t)^{-6}\). This is a much more rapid decrease than for radiation (\(\rho \propto a^{-4}\)) or matter (\(\rho \propto a^{-3}\)) \cite{Chavanis2015}. The critical point $H_{D}$ is corresponding to a non-accelerating, DE-dominated Universe with
 a stiff DE. In this case, the EoS parameter $\omega_{tot}=1$. This critical point exists in studying the standard quintessence model and the teleparallel DE model \cite{copelandLiddle, Xu_2012}.
\end{itemize}

\textbf{The Eigenvalues and the Stability Conditions :}

\begin{itemize}
\item \textbf{Stability of Critical Points} $A_{R}, B_{R}:$ \\
The eigenvalues of the critical points $A_R$ are as 
$\left[\nu_{1}=2, \, \nu_{2}=-1,\, \nu_{3}=1, \, \nu_{4}=0\right].$ According to the sign of these eigenvalues, one can conclude that this critical point is a saddle point. The eigenvalues of the critical points $B_R$ are as $\left[\nu_{1}=-\frac{4 \alpha }{\lambda }, \, \nu_{2}=1, \, \nu_{3}=-\frac{\sqrt{64 \lambda ^4-15 \lambda ^6}}{2 \lambda ^3}-\frac{1}{2}, \, \nu_{4}=\frac{\sqrt{64 \lambda ^4-15 \lambda ^6}}{2 \lambda ^3}-\frac{1}{2}\right].$ This critical point is not an unstable critical point but is showing the saddle behaviour at $\left[0<\lambda <2\land \alpha <0\right].$ Both of these critical points represent the radiation-dominated epoch of the evolution of the Universe.\\

\item \textbf{Stability of Critical Points} $C_{M}, D_{M}:$ \\
The eigenvalues of the critical point $C_{M}$ are $\left[\nu_{1}=-\frac{3}{2} ,\, \nu_{2}=\frac{3}{2} ,\, \nu_{3}=-\frac{1}{2} ,\, \nu_{4}=0 ,\, \right]$. The existence of the positive and the negative eigenvalue implies that this critical point is a saddle critical point. The eigenvalues at the critical point $D_{M}$ are $\left[\nu_{1}=-\frac{3 \alpha }{\lambda } ,\, \nu_{2}=-\frac{1}{2} ,\, \nu_{3}=-\frac{3 \left(\lambda ^3+\sqrt{24 \lambda ^4-7 \lambda ^6}\right)}{4 \lambda ^3} ,\, \nu_{4}=\frac{3 \sqrt{24 \lambda ^4-7 \lambda ^6}}{4 \lambda ^3}-\frac{3}{4} \right]$. It is stable at $\left[\left(\alpha <0\land -2 \sqrt{\frac{6}{7}}\leq \lambda <-\sqrt{3}\right)\lor \left(\alpha >0\land \sqrt{3}<\lambda \leq 2 \sqrt{\frac{6}{7}}\right)\right]$, and saddle at \\
$\left[\left(\alpha <0\land \sqrt{3}<\lambda \leq 2 \sqrt{\frac{6}{7}}\right)\lor \left(\alpha >0\land -2 \sqrt{\frac{6}{7}}\leq \lambda <-\sqrt{3}\right)\right]$.\\
The saddle point nature of these critical points represent early-time matter-dominated epochs of the evolution of the Universe as expected.\\
\item \textbf{Stability of Critical Points} $E_{DE}, F_{DE}, G_{DE}:$ \\
The eigenvalues at the Jacobian matrix for a critical point $E_{DE}$ are $\Big[\nu_{1}=-\alpha  \lambda,\, \nu_{2}=\frac{1}{2} \left(\lambda ^2-6\right),\, \nu_{3}=\frac{1}{2} \left(\lambda ^2-4\right),\, \nu_{4}=\lambda ^2-3 \Big]$. This critical point is a stable late-time attractor solution with stability at $\Big(-\sqrt{3}<\lambda <0\land \alpha <0\Big)\lor \left(0<\lambda <\sqrt{3}\land \alpha >0\right)$ and is saddle at $\left(-\sqrt{3}<\lambda <0\land \alpha >0\right)\lor \left(0<\lambda <\sqrt{3}\land \alpha <0\right)$. This critical point is unstable for $\Big(\lambda <-\sqrt{6}\land \alpha >0\Big)\lor \Big(\lambda >\sqrt{6}\land \alpha <0\Big)$. The critical point $F_{DE}$ represents the cosmological constant (de-Sitter) solution with $\omega_{tot}=-1$. The eigenvalues at this critical point are $\Big[\nu_{1}=-3,\, \nu_{2}=-2,\, \nu_{3}=-\frac{3 \left(\sqrt{\left(\lambda ^2+6\right) \left(8 \alpha  \lambda +\lambda ^2+6\right)}+\lambda ^2+6\right)}{2 \left(\lambda ^2+6\right)},\, \nu_{4}=\frac{3}{2} \left(\frac{\sqrt{\left(\lambda ^2+6\right) \left(8 \alpha  \lambda +\lambda ^2+6\right)}}{\lambda ^2+6}-1\right)\Big]$. It shows stability at $\Big[8 \alpha  \lambda +\lambda ^2+6\geq 0\land ((\lambda >0\land \alpha <0)\lor (\alpha >0\land \lambda <0))\Big]$ and saddle at $\Big[ (\lambda <0\land \alpha <0)\lor (\lambda >0\land \alpha >0)\Big]$. The eigenvalues at the critical point $G_{DE}$ are $\Big[\nu_{1}=-2,\, \nu_{2}=0,\, \nu_{3}=-3,\,\nu_{4}=-3\Big]$. Note that the critical
point with zero eigenvalues is termed a non-hyperbolic critical point. This is the critical point containing a zero eigenvalue, and the other three eigenvalues are negative; hence, to check its stability, the linear stability theory fails to obtain the stability of such a critical point. Hence, We have obtained stability at this critical point using the Central Manifold Theory (CMT), and the detailed application of CMT is presented in the \textbf{Appendix-}\ref{CMT}. From CMT, It has been concluded that this critical point is stable at $\alpha>0$. \\ 

\item \textbf{Stability of Critical Point} $H_{D}:$ \\
The eigenvalues at this critical point are $\left[\nu_{1}=3,\,\nu_{2}=1,\,\nu_{3}=-\sqrt{6} \alpha,\,\nu_{4}=3-\sqrt{\frac{3}{2}} \lambda\right]$. This critical point is saddle at $\left(\alpha >0\land \lambda >\sqrt{6}\right)$ and is unstable at $\left(\alpha <0\land \lambda <\sqrt{6}\right)$.

\end{itemize}

\textbf{Numerical Results:}\\

In this study, we have analysed critical points representing different epochs of the evolution of Universe. Among these critical points, $A_{R}, B_{R}$ are the critical points describing the radiation-dominated epoch of the evolution of the Universe. Moreover, these critical points show saddle point behaviour. The critical points $C_{M}, D_{M}$ represent the matter-dominated epoch of the Universe, and both show saddle point behaviour. The critical points $E_{DE}, F_{DE}$ and $G_{DE}$ are the DE- solutions. From the stability analysis, one can confirm that these critical points are the late time stable attractors within the particular range of the model parameters. Now, we will analyse the numerical solution using the autonomous system presented in Eq. \eqref{dynamicalsystem}. The numerical solutions were calculated using the ND-solve command in Mathematica. Our analysis was based on the Hubble and Supernovae Ia (SNe Ia) observational data sets, presented in detail in {\bf Appendix-\ref{Datasets}}. \\

\begin{figure}[H]
\centering
\begin{subfigure}[h]{0.3\textwidth}
\centering
\includegraphics[width=78mm]{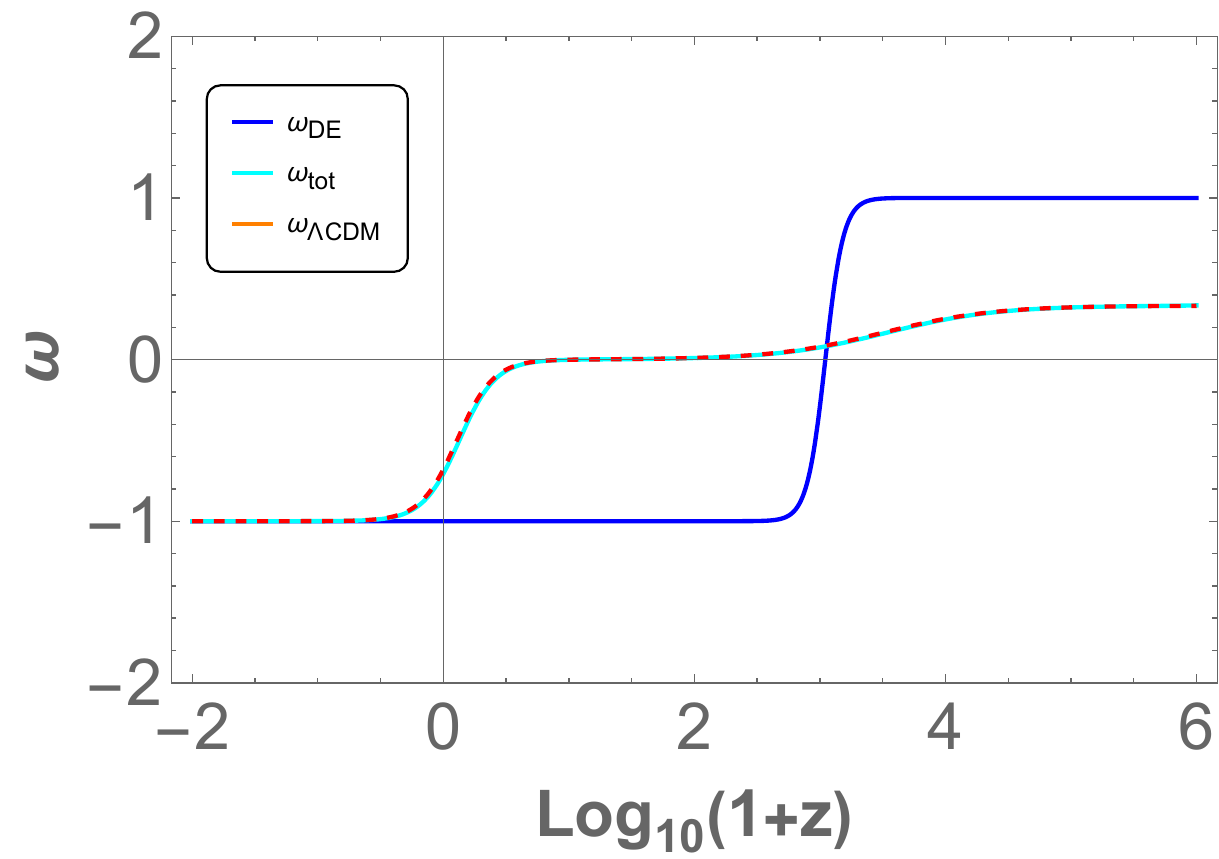}
\caption{The behaviour of EoS parameters.}\label{Eosm1}
\end{subfigure}
\hspace{1.9cm}
\begin{subfigure}[h]{0.5\textwidth}
\centering
\includegraphics[width=78mm]{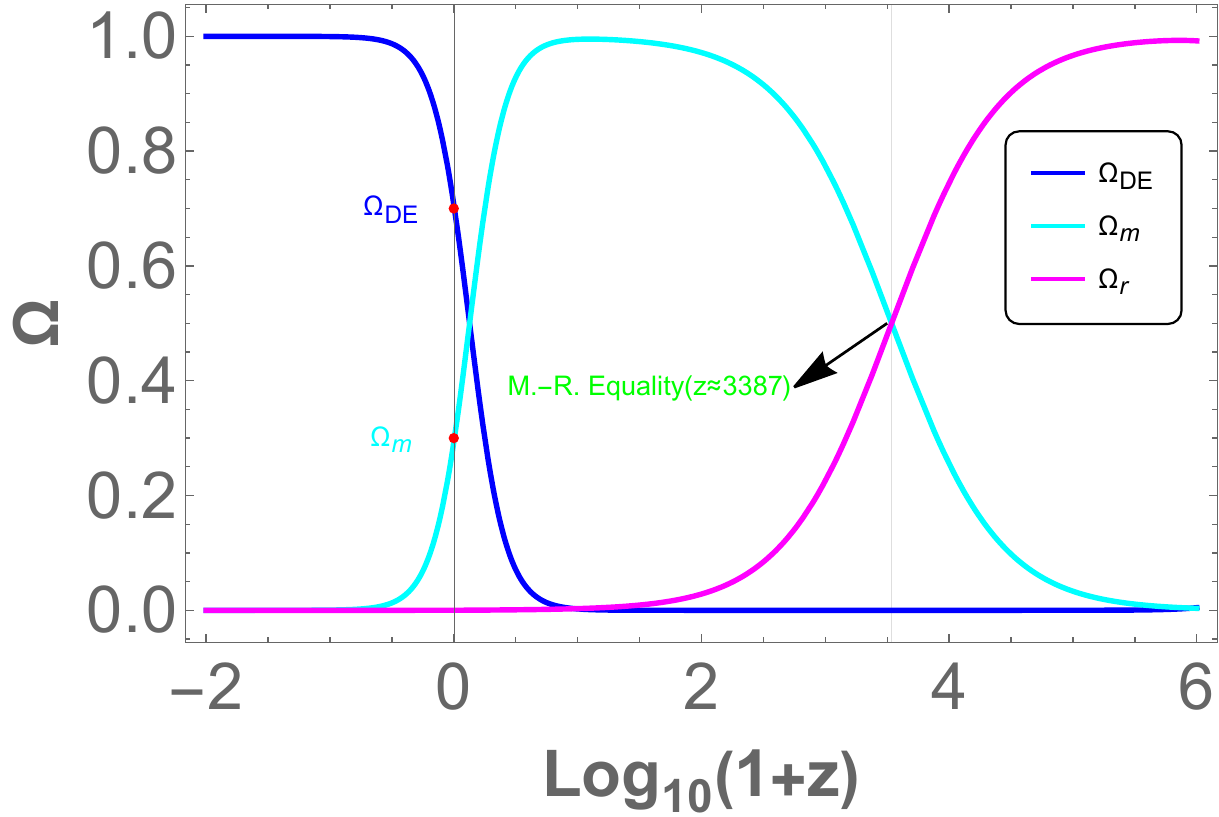}
\caption{The behaviour of standard density parameters.}\label{densityparametersm1}
\end{subfigure}
\caption{In this case, the initial conditions are $x_C=10^{-8.89} ,\,y_C=10^{-2.89} ,\,u_C=10^{-5.96} ,\,\rho_C=10^{-0.75} ,\, \lambda=-0.01, \alpha=-5.2$. }\label{Eosdensitym1}
\end{figure}

\begin{figure}[H]
\centering
\begin{subfigure}[h]{0.35\textwidth}
\centering
\includegraphics[width=78mm]{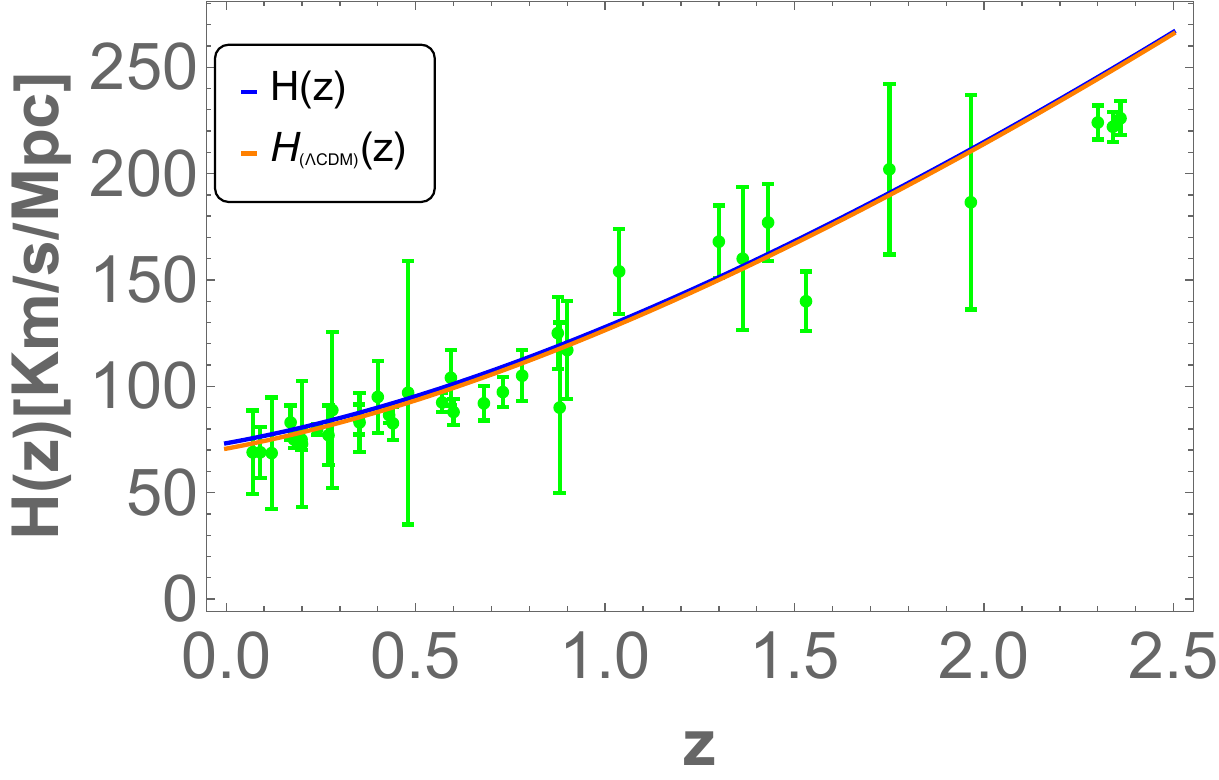}
\caption{The behaviour of Hubble parameter.} \label{H(z)m1}
\end{subfigure}
\hspace{2.1cm}
\begin{subfigure}[h]{0.35\textwidth}
\centering
\includegraphics[width=78mm]{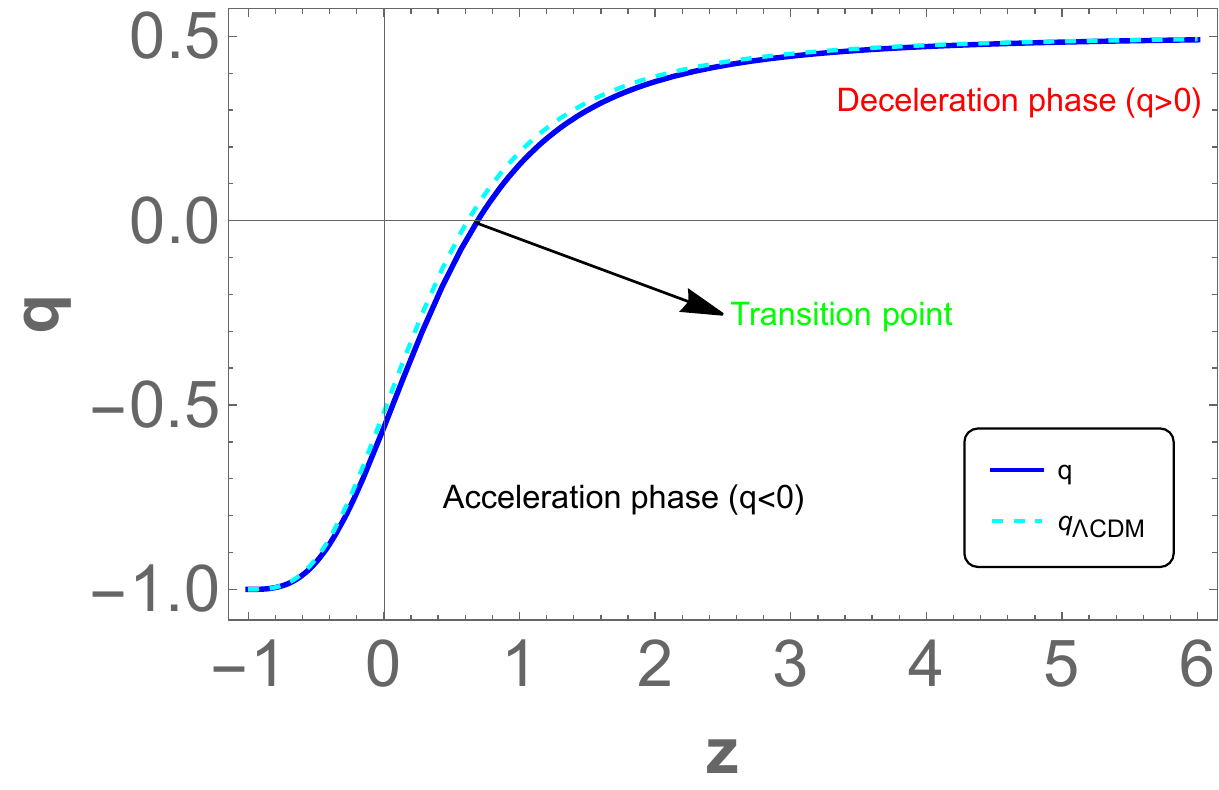}
\caption{The behaviour of deceleration parameter.}\label{decelerationparametersm1}
\end{subfigure}
\caption{The initial conditions are: $x_C=10^{-8.89} ,\,y_C=10^{-2.89} ,\,u_C=10^{-5.96} ,\,\rho_C=10^{-0.75} ,\, \lambda=-0.01, \alpha=-5.2$. }\label{h(z)q(z)m1}
\end{figure}

\begin{figure}[H]
    \centering
\includegraphics[width=78mm]{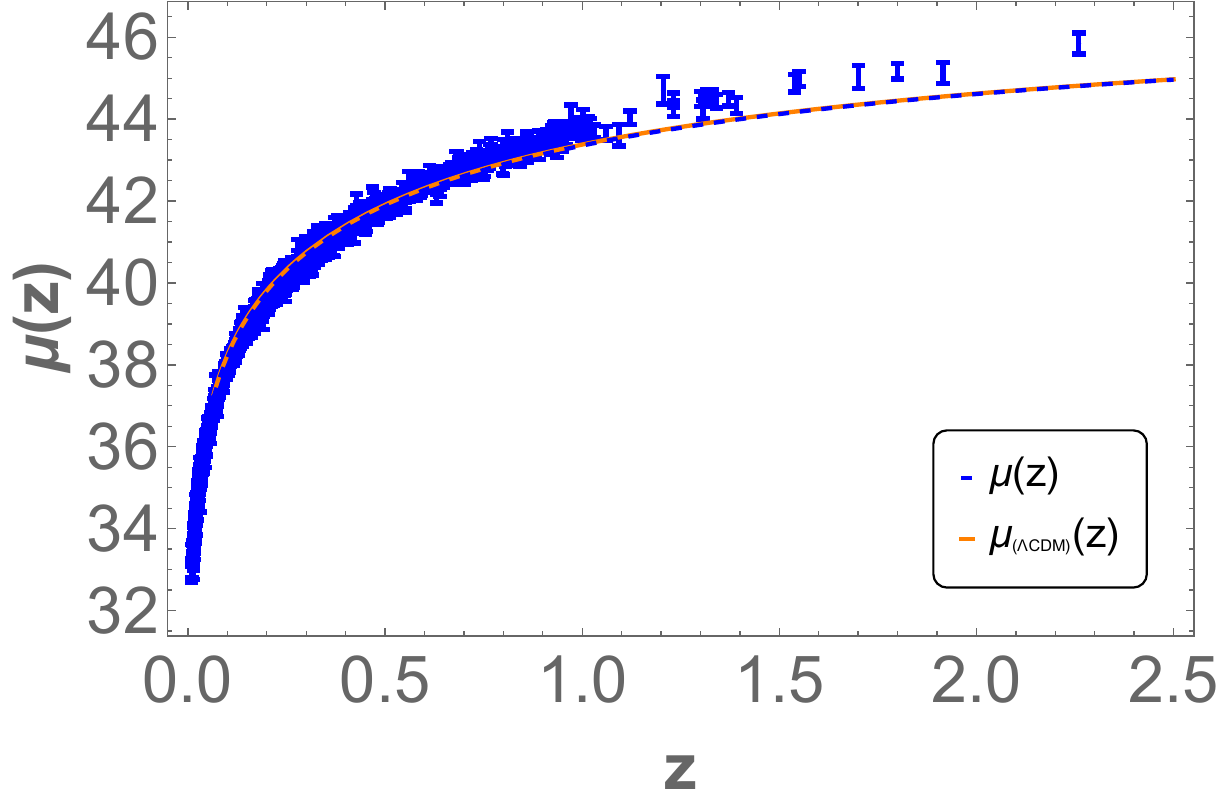}\caption{Plot of the observed distance modulus function $\mu(z)$ and the predicted $\Lambda$CDM model distance modulus function $\mu_{\Lambda CDM}(z)$. The initial conditions are: $x_C=10^{-8.89}$, $y_C=10^{-2.89}$, $u_C=10^{-5.96}$, $\rho_C=10^{-0.75}$, $\lambda=-0.01$, and $\alpha=-5.2$. } \label{mu(z)m1}
\end{figure}

In Figure \ref{Eosm1}, the EoS parameters for DE, total, and the EoS parameter of $\Lambda$CDM are compared. The plots demonstrate the evolution of $\omega_{tot}$, $\omega_{DE}$, and $\omega_{\Lambda CDM}$ towards $-1$ at late times. The present value of $\omega_{DE}(z=0)=-1$, consistent with the observational result of Planck Collaboration \cite{Aghanim:2018eyx}. The evolution of energy densities of radiation, DE, and DM is illustrated in Figure \ref{densityparametersm1}. From these plots, one can observe that radiation prevails in the early times of the Universe. Moreover, the plots describe the sequence by showing that the DM-dominated era can be described as a small time period and, finally, the emergence of the cosmological constant at the late time is observed. The plot shows the contribution amount of DM $\Omega_{m}\approx 0.3$ and DE $\Omega_{DE}\approx 0.7$ density parameters. The time of matter-radiation equality is around $z\approx 3387$ and is denoted with a pointed arrow in Figure \ref{densityparametersm1}. Figure \ref{H(z)m1} illustrates the Hubble rate evolution and the Hubble data points \cite{Moresco_2022_25}, with $H_{0}=70$ Km/(Mpc sec) \cite{Aghanim:2018eyx}, showing that the model closely aligns with the standard $\Lambda$CDM model. In Figure \ref{decelerationparametersm1}, the behavior of the deceleration parameter is examined; it shows transient behavior at $z\approx 0.66$ and is consistent \cite{PhysRevD.90.044016a}. At present time, the value of the deceleration parameter is $q\approx -0.53$ \cite{PhysRevResearch.2.013028}. Figure \ref{mu(z)m1} illustrates the evolution of the modulus function $\mu(z)$, showing that the model curve aligns well with the $\Lambda$CDM model modulus function $\mu_{\Lambda CDM}$ including data from 1048 Supernovae Ia (SNe Ia).\\

\begin{figure}[H]
    \centering
\includegraphics[width=78mm]{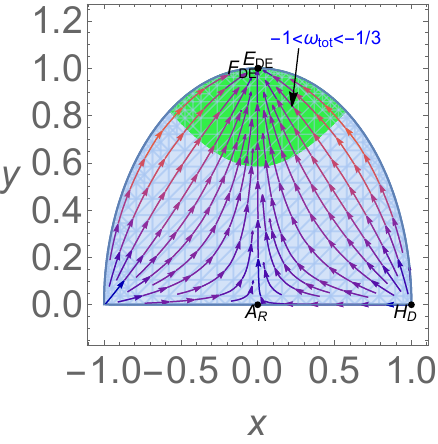}\caption{Visualization of the 2D phase space dynamics with initial conditions and parameter values consistent with Figure \ref{Eosdensitym1}. The accelerating (quintessence) phase of the Universe, $-1 < \omega_{tot} < -\frac{1}{3}$, is highlighted in the green/shaded region. } \label{2dm1}
\end{figure}

In the 2-D phase space portrait shown in Figure \ref{2dm1}, where the region is plotted by using the conditions $0<\Omega_{m}\leq1$ and $y>0$. The green-colored shaded region corresponds to the accelerating expansion of the Universe. Specifically it describes the quintessence behavior ($-1<\omega_{tot}<-\frac{1}{3}$), with the stable critical points $E_{DE}$ and $F_{DE}$ lying within this region. These critical points characterize the late-time cosmic acceleration phase of the Universe.

The dynamical variables $x$ and $y$ are used to plot the phase portrait. The critical points $A_{R}$ and $E_{DE}$ are connected by the trajectory showing a red line. The phase space trajectories indicate that the solution transitions from a saddle critical point ($A_{R}$) to a stable critical point ($E_{DE}, F_{DE}$). The behavior of the trajectories suggests that the critical points $A_{R}$, $H_{D}$, $D_{M}$, and $B_{R}$ exhibit saddle behavior, while critical points $E_{DE}$ and $F_{DE}$ demonstrate stable behavior.  The initial conditions and the values of the model parameters are consistent with those of Figure \ref{Eosdensitym1}.

\subsection{Potential \texorpdfstring{$P_2 \,: Cosh(\xi \phi)-1$}{}}\label{P_2}

This potential plays an important role in studying the dynamics of the phantom and the quintessence DE, DM models \cite{Roy_2018,Sahni_2000}. The critical points are presented in Table \ref{P2criticalpoints}. In this case, from Table \ref{Potentialfunctions}, the value of $\alpha_1=-\frac{1}{2}$ and $\alpha_3=\frac{\xi^2}{2}$, hence we will analyse this potential with the autonomous dynamical system having five independent variables. 
\begin{table}[H]
    \centering 
    \begin{tabular}{|c |c |c |c| c| c| c|} 
    \hline\hline 
    \parbox[c][0.9cm]{1.3cm}{\textbf{C. P.}
    }& $ \{ x, \, y, \, u, \, \rho, \, \lambda \} $ & $q$ &  \textbf{$\omega_{tot}$}& $\Omega_{r}$& $\Omega_{m}$& $\Omega_{DE}$\\ [0.5ex] 
    \hline\hline 
    \parbox[c][1.3cm]{1.3cm}{$a_{R}$ } &$\{ 0, 0, 0,  1,  \xi \}$ & $1$ &  $\frac{1}{3}$& $1$& $0$ & $0$ \\
    \hline
    \parbox[c][1.3cm]{1.3cm}{$b_{R}$ } & $\left\{\frac{2 \sqrt{\frac{2}{3}}}{\xi },\frac{2}{\sqrt{3} \xi },0,\frac{\sqrt{\xi ^2-4}}{\xi },\xi \right\}$ & $1$&  $\frac{1}{3}$& $1-\frac{4}{\xi ^2}$ & $0$ & $\frac{4}{\xi ^2}$ \\
    \hline
   \parbox[c][1.3cm]{1.3cm}{$c_{M}$ } &  $\{0,\, \, 0,\, \, 0, \, \, 0, \, \, \xi\}$ & $\frac{1}{2}$ &  $0$& $0$ & $1$ & $0$\\
   \hline
   \parbox[c][1.3cm]{1.3cm}{$d_{M}$} &   $\left\{\frac{\sqrt{\frac{3}{2}}}{\xi },\frac{\sqrt{\frac{3}{2}}}{\xi },0,0,\xi\right\}$ & $\frac{1}{2}$ &  $0$& $0$ & $1-\frac{3}{\xi ^2}$ & $\frac{3}{\xi ^2}$\\
   \hline
 \parbox[c][1.3cm]{1.3cm}{$e_{DE}$} &  $\left\{\frac{\xi }{\sqrt{6}},\sqrt{1-\frac{\xi ^2}{6}},0,0,\xi\right\}$ & $-1+\frac{\xi ^2}{2}$ & $-1+\frac{\xi ^2}{3}$ & $0$ & $0$ & $1$\\
 \hline
 \parbox[c][1.3cm]{1.3cm}{$f_{DE}$} &  $\left\{0,1,\frac{\xi }{3},0,\xi\right\}$ & $-1$ &  $-1$ & $0$ & $0$ & $1$\\
 \hline
 \parbox[c][1.3cm]{1.3cm}{$g_{DE}$} &  $\left\{ 0,1,0,0,0\right\}$ & $-1$ &  $-1$ & $0$ & $0$ & $1$\\
 \hline
 \parbox[c][1.3cm]{1.3cm}{$h_{D}$} &  $\left\{1,0,0,0,\xi\right\}$ & $2$ &  $1$ & $0$ & $0$ & $1$\\
 \hline
    \end{tabular}
    \caption{Critical points analysis: values of $q$,\, $\omega_{tot}$,\, $\Omega_{r}$,\, $\Omega_{m}$ 
 and $\Omega_{DE}$ for potential \ref{P_2} $(P_2)$.}
  \label{P2criticalpoints}
\end{table}
 In this potential function, one thing can be noticed in most of the critical points, the dynamical variable $\lambda$ equals the potential parameter $\xi$. The detailed analysis of the critical point is presented as follows,
\begin{itemize}

\item{\textbf{Radiation-Dominated Critical Points} $a_R, b_R$}: \\
The critical point $a_{R}$ is a standard radiation-dominated critical point with $\Omega_{r}=1$. This critical point describes the early phase of the evolution of the Universe with $\omega_{tot}=\frac{1}{3}$. The standard density parameter for the radiation $\Omega_{r}$ takes the value 1 at this critical point. The critical point $b_{R}$ describes the non-standard radiation-dominated era with $\Omega_{r}=1-\frac{4}{\xi^2}$. This critical point is a scaling solution, and the physical condition on the density parameter $0<\Omega_{DE}<1$ applies the condition on parameter $\xi$, such that $|\xi>2|$. These critical points are considered in \cite{Gonzalez-Espinoza:2020jss} but are not discussed in \cite{copelandLiddle, Xu_2012}.

\item{\textbf{Matter-Dominated Critical Points} $c_M, d_M$}: \\
The critical point $c_{M}$ is the standard matter dominated critical point with $\Omega_{M}=1$ at which the value of $q=\frac{1}{2}$ and $\omega_{tot}=0$. Another critical point representing the cold DM-dominated era in this case is $d_M$. This critical point is the scaling matter-dominated solution with $\Omega_{DE}=\frac{3}{\xi^2}$. The condition on $0<\Omega_{DE}<1$, applies the condition on the parameter $\xi$ as $|\xi|>\sqrt{3}.$ The critical point $d_{M}$ appeared in the study made in the standard quintessence model and the teleparallel DE model \cite{Xu_2012,copelandLiddle}.

\item{\textbf{DE-Dominated Critical Points} $e_{DE}, f_{DE}, g_{DE}$}: \\
The critical point $e_{DE}$ is the critical point representing the DE-dominated era of the evolution of the Universe. The value of $\omega_{tot}=-1+\frac{\xi^2}{3}$. This critical point describes the late time cosmic acceleration at $-\sqrt{2}<\xi <0\lor 0<\xi <\sqrt{2}$ and a late-time attractor solution. The points $f_{DE}, g_{DE}$ are the cosmological constant solutions with the value of $\omega_{tot}=-1$. So, $e_{DE}$, $f_{DE}$ and $g_{DE}$ represent the standard DE era of the Universe evolution with $\Omega_{DE}=1$.

 \item{\textbf{Critical Point Representing the Stiff DE $h_D$ :}} 
 This critical point describes the stiff matter-dominated era. During the stiff matter-dominated era, the Universe would expand and cool rapidly. The energy density of stiff matter decreases rapidly with the expansion compared to other forms of matter or radiation.  

\end{itemize}

\textbf{The Eigenvalues and the Stability Conditions :}

\begin{itemize}
\item \textbf{Stability of Critical Points} $a_{R}, b_{R}:$ \\
The eigenvalues at this critical point are
$\Big[\nu_{1}=2,\, \nu_{2}=-1,\, \nu_{3}=1,\, \nu_{4}=0,\,\nu_{5}=0 \Big]$. The presence of both positive and negative eigenvalues imply the saddle behaviour. The eigenvalues at $b_{R}$ are $\Big[\nu_{1}=-\frac{4 \alpha }{\xi },\, \nu_{2}=1, \, \nu_{3}=4, \, \nu_{4}=-\frac{\sqrt{64 \xi ^4-15 \xi ^6}}{2 \xi ^3}-\frac{1}{2}, \, \nu_{5}=\frac{\sqrt{64 \xi ^4-15 \xi ^6}}{2 \xi ^3}-\frac{1}{2}\Big]$. All the eigenvalues will not take the positive value; this critical point is not an unstable critical point and is always shows the saddle point behaviour at $\left[\alpha <0\land -2<\xi <0\right]$. \\

\item \textbf{Stability of Critical Points} $c_{M}, d_{M}:$ \\
The eigenvalues at the point $c_{M}$ are $\Big[\nu_{1}=-\frac{3}{2},\,\nu_{2}=\frac{3}{2},\,\nu_{3}=-\frac{1}{2},\,\nu_{4}=0,\,\nu_{5}=0,\,\Big]$, shows the saddle characteristics. The corresponding eigenvalues can be written as, $\Big[\nu_{1}=-\frac{3 \alpha }{\xi },\,\nu_{2}=-\frac{1}{2},\,\nu_{3}=3,\,\nu_{4}=-\frac{3 \left(\xi ^3+\sqrt{24 \xi ^4-7 \xi ^6}\right)}{4 \xi ^3},\,\nu_{5}=\frac{3 \sqrt{24 \xi ^4-7 \xi ^6}}{4 \xi ^3}-\frac{3}{4}\Big]$. This critical point is saddle at $\left[\alpha <0\land -\sqrt{3}<\xi <0\right]$.

\item \textbf{Stability of Critical Points} $e_{DE}, f_{DE}, g_{DE}:$ \\
The eigenvalues of critical point $e_{DE}$ are $\Big[\nu_{1}=-\alpha  \xi,\, \nu_{2}=\xi ^2,\, \nu_{3}=\frac{1}{2} \left(\xi ^2-6\right),\, \nu_{4}=\frac{1}{2} \left(\xi ^2-4\right),\, \nu_{5}= ,\xi ^2-3\Big]$ and it shows stability at $\left(\alpha <0\land -\sqrt{3}<\xi <0\right)\lor \left(\alpha >0\land 0<\xi <\sqrt{3}\right)$ and is saddle at$\left(\alpha <0\land 0<\xi <\sqrt{3}\right)\lor \left(\alpha >0\land -\sqrt{3}<\xi <0\right)$ and is unstable at $\left(\alpha <0\land \xi >\sqrt{6}\right)\lor \left(\alpha >0\land \xi <-\sqrt{6}\right)$.\\ The eigenvalues at the critical point $f_{DE}$ are, $\Big[\nu_{1}=0,\, \nu_{2}=-3,\, \nu_{3}=-2,\, \nu_{4}=-\frac{3 \left(\sqrt{\left(\xi ^2+6\right) \left(8 \alpha  \xi +\xi ^2+6\right)}+\xi ^2+6\right)}{2 \left(\xi ^2+6\right)},\, \nu_{5}= \frac{3}{2} \left(\frac{\sqrt{\left(\xi ^2+6\right) \left(8 \alpha  \xi +\xi ^2+6\right)}}{\xi ^2+6}-1\right)\Big]$. We have obtained the stability at this critical point using the CMT, and this critical point is stable at $(\lambda<0\land \xi >0)\lor (\lambda>0\land \xi >0)$ [Ref. \textbf{Appendix}-\ref{CMT}].\\

The eigenvalues at critical point $g_{DE}$ are $\Big[\nu_{1}=-3,\, \nu_{2}=-2,\, \nu_{3}=0,\, \nu_{4}=\frac{1}{2} \left(-\sqrt{9-6 \xi ^2}-3\right),\, \nu_{5}=\frac{1}{2} \left(\sqrt{9-6 \xi ^2}-3\right) \Big]$. According to the CMT, this critical point shows stable behaviour for  $\alpha>0$ [Ref. \textbf{Appendix}-\ref{CMT}]. \\
\item \textbf{Stability of Critical Point} $h_{D}:$ \\
The eigenvalues at the critical point $h_{D}$ are $\Big[\nu_{1}=3,\, \nu_{2}=1,\, \nu_{3}=-\sqrt{6} \alpha,\, \nu_{4}=\sqrt{6} \xi,\, \nu_{5}= 3-\sqrt{\frac{3}{2}} \xi\Big]$. This critical point is saddle at $\left[\alpha >0\land \xi >\sqrt{6}\right]$ and is unstable at $\left[\alpha <0\land \xi <\sqrt{6}\right]$. 

\end{itemize}

\textbf{Numerical Results:}

\begin{figure}[H]
\centering
\begin{subfigure}[h]{0.3\textwidth}
\centering
\includegraphics[width=78mm]{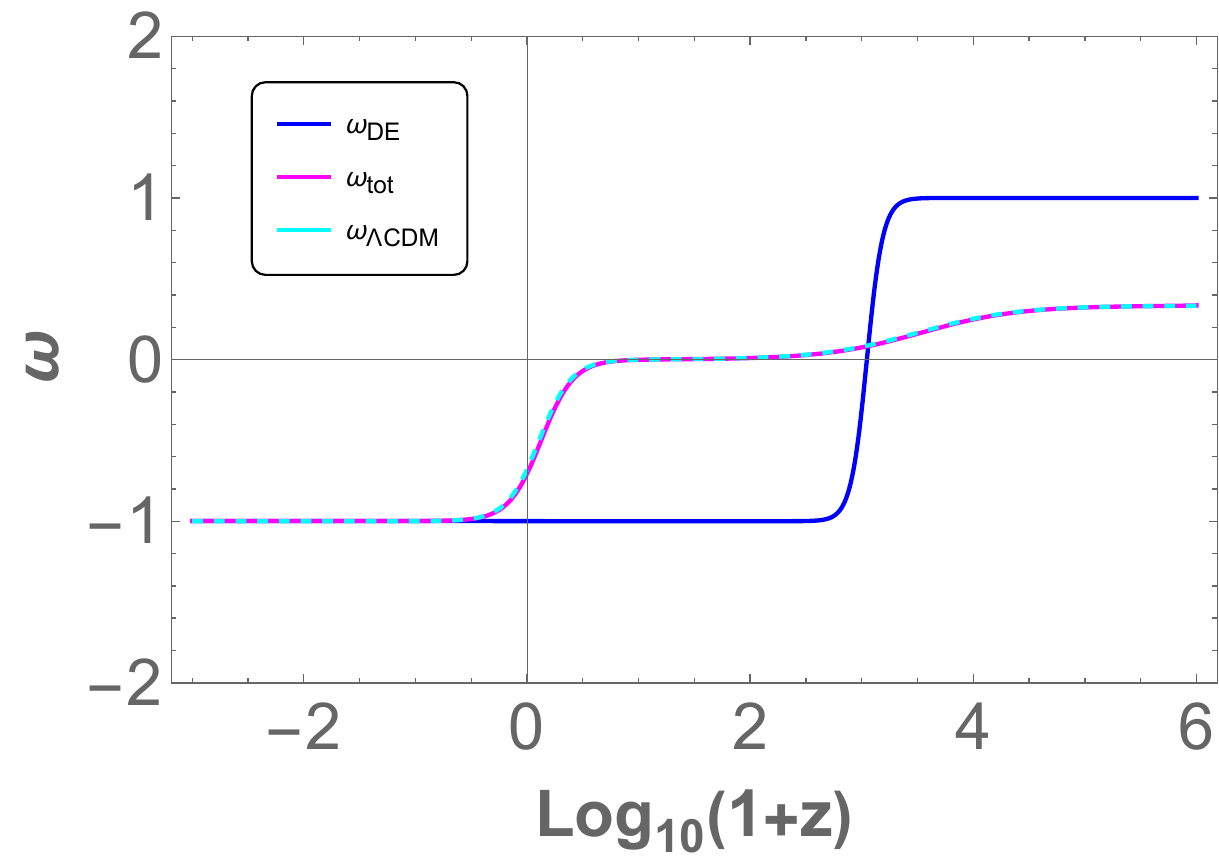}
\caption{The behaviour of EoS parameters.}\label{Eosm2}
\end{subfigure}
\hspace{1.9cm}
\begin{subfigure}[h]{0.5\textwidth}
\centering
\includegraphics[width=78mm]{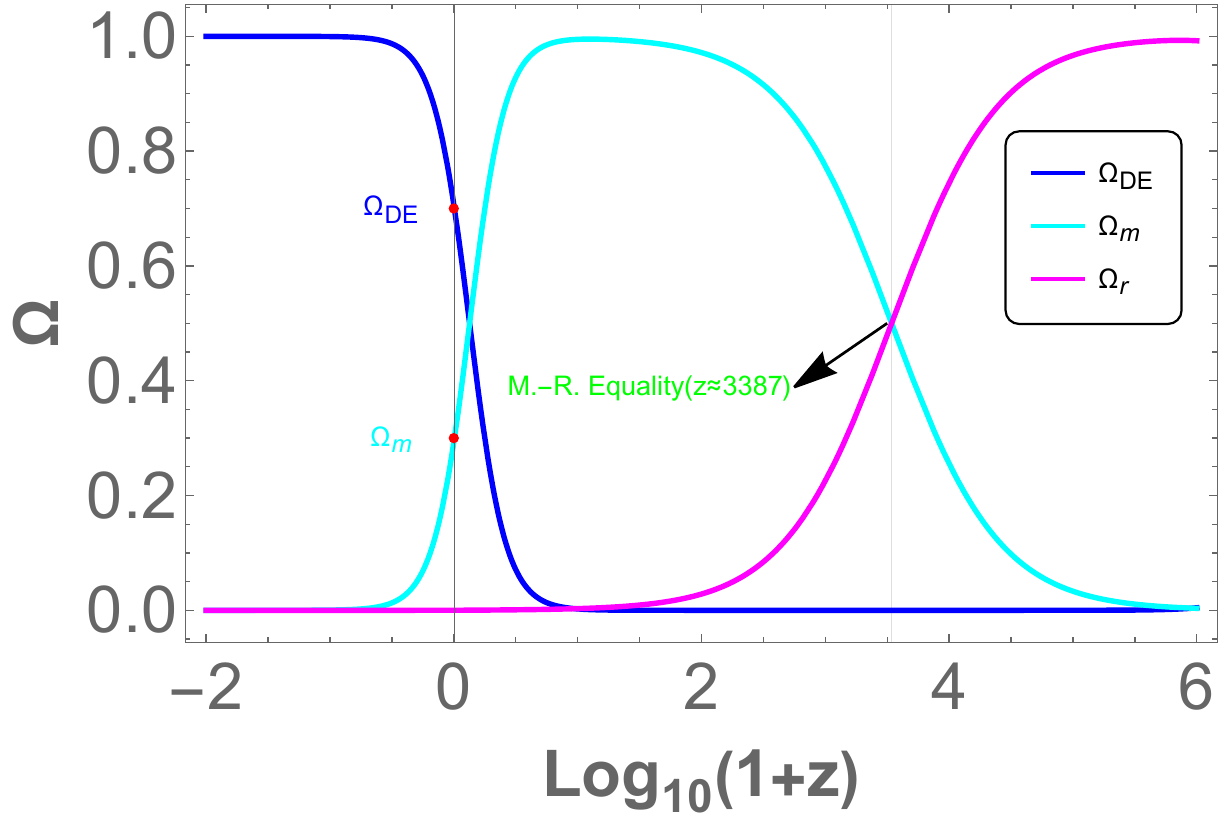}
\caption{The behaviour of standard density parameters.}\label{densityparametersm2}
\end{subfigure}
\caption{The initial conditions are: $x_C=10^{-8.89} ,\,y_C=10^{-2.89} ,\,u_C=10^{-5.96} ,\,\rho_C=10^{-0.75}, \lambda_{c}=10^{-1.3}, \, \alpha=-5.2, \, \xi= -0.02$. }\label{Eosdensitym2}
\end{figure}

\begin{figure}[H]
\centering
\begin{subfigure}[h]{0.35\textwidth}
\centering
\includegraphics[width=78mm]{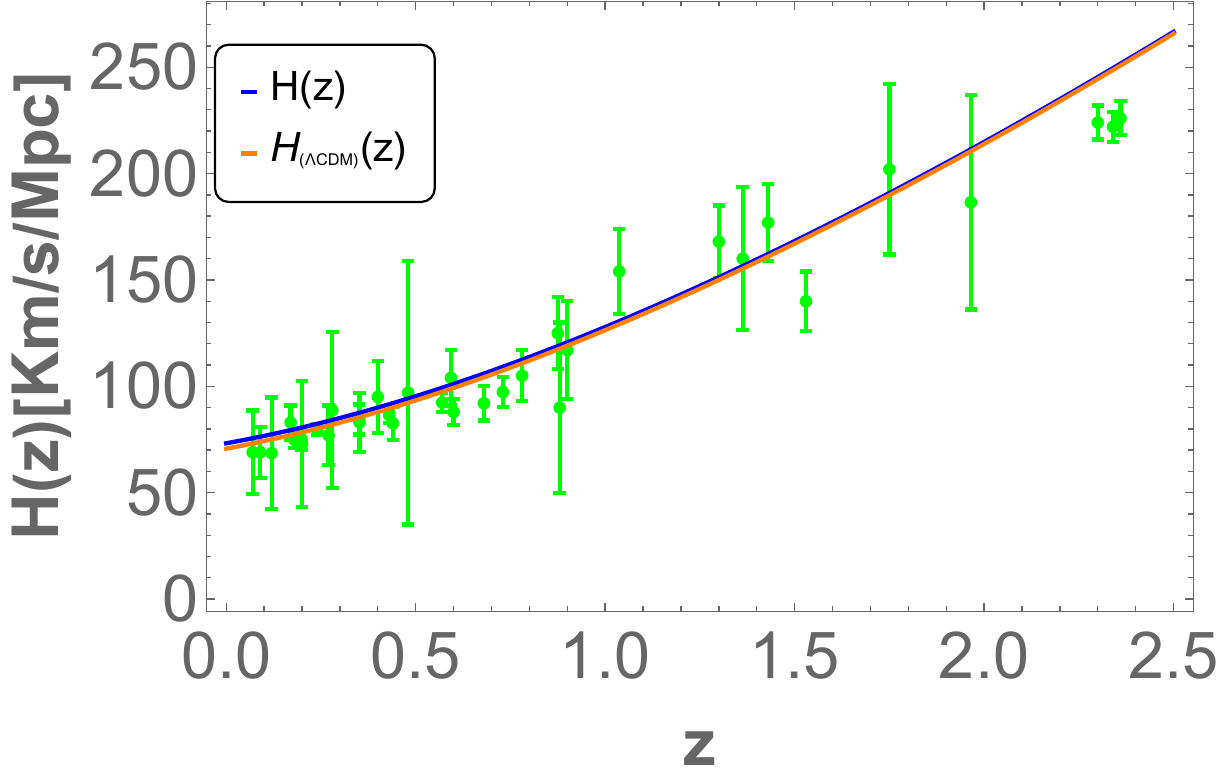}
\caption{The behaviour of Hubble parameter.}\label{H(z)m2}
\end{subfigure}
\hspace{1.9cm}
\begin{subfigure}[h]{0.35\textwidth}
\centering
\includegraphics[width=78mm]{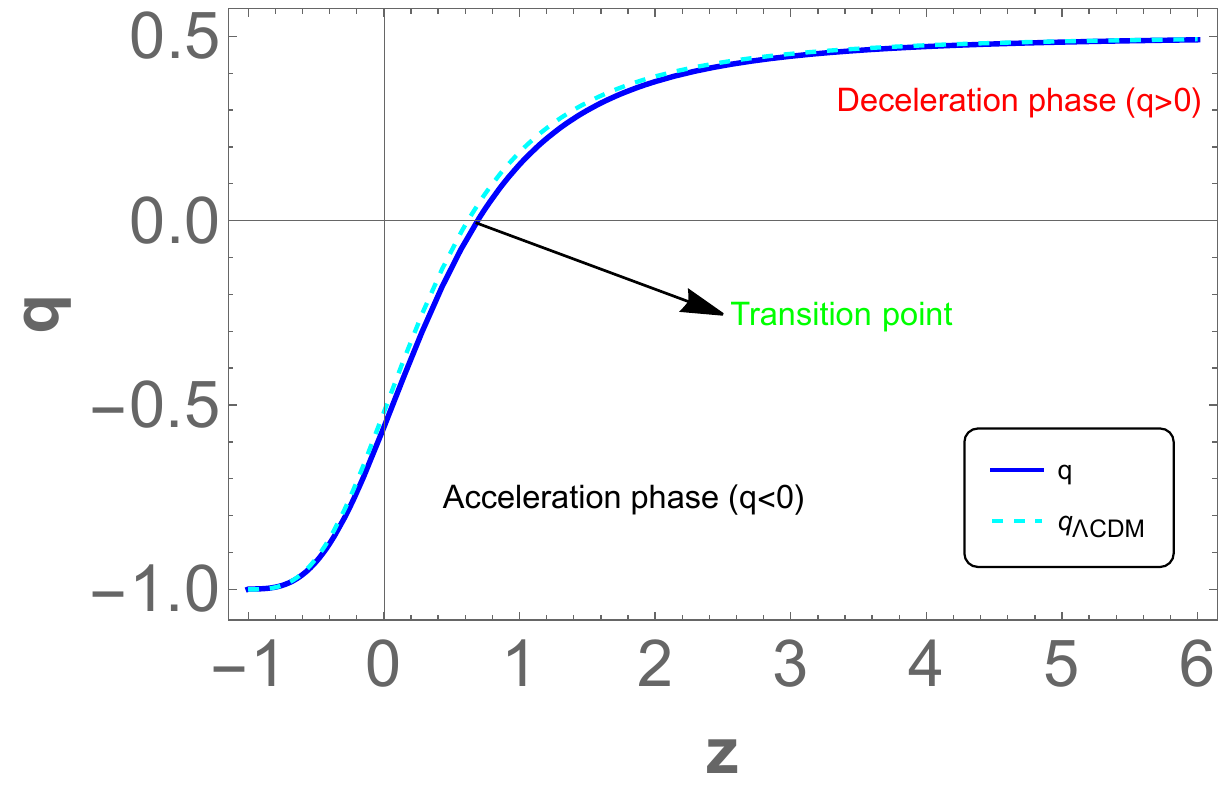}
\caption{The behaviour of deceleration parameter.}\label{decelerationparametersm2}
\end{subfigure}
\caption{The initial conditions are: $x_C=10^{-8.89} ,\,y_C=10^{-2.89} ,\,u_C=10^{-5.96} ,\,\rho_C=10^{-0.75}, \lambda_{c}=10^{-1.3}, \, \alpha=-5.2, \, \xi= -0.02$. }\label{h(z)q(z)m2}
\end{figure}

\begin{figure}[H]
    \centering
\includegraphics[width=78mm]{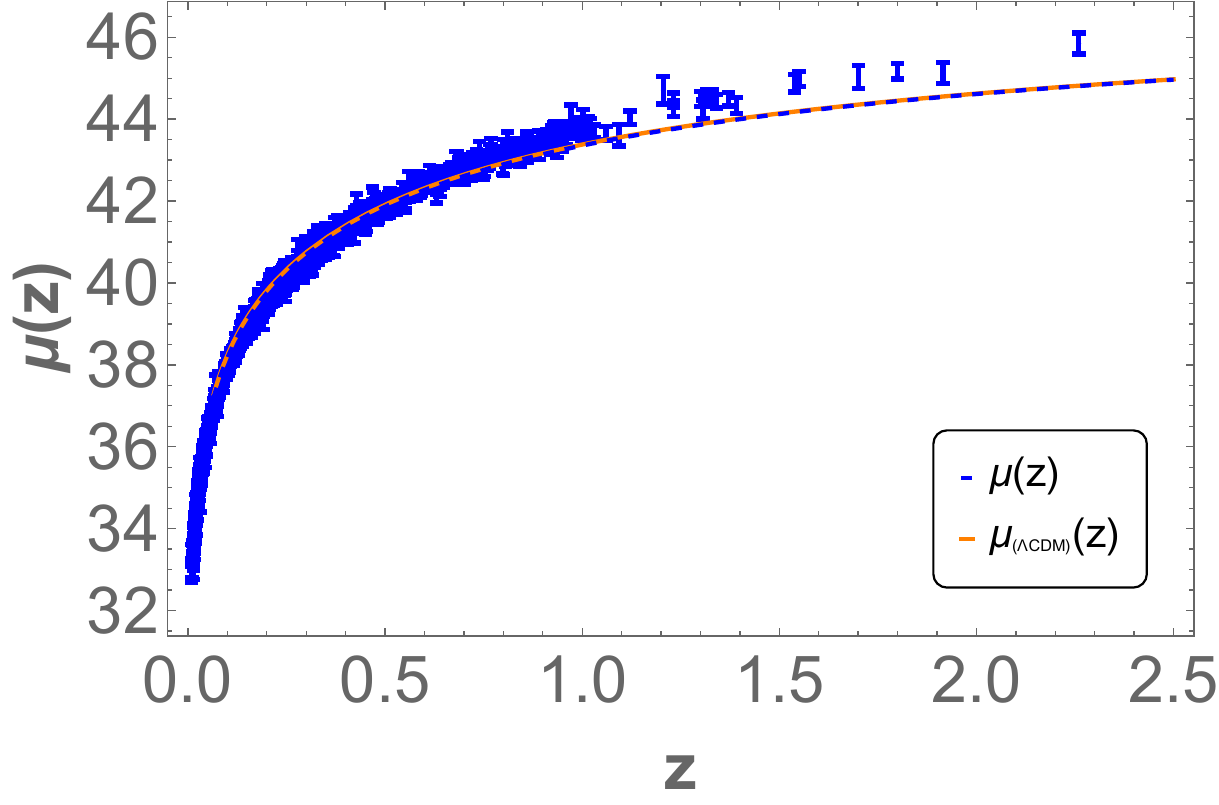}\caption{Plot of the observed distance modulus function $\mu(z)$ and the predicted $\Lambda$CDM model distance modulus function $\mu_{\Lambda CDM}(z)$. The initial conditions are:  $x_C=10^{-8.89} ,\,y_C=10^{-2.89} ,\,u_C=10^{-5.96} ,\,\rho_C=10^{-0.75}, \lambda_{c}=10^{-1.3}, \, \alpha=-5.2, \, \xi= -0.02$. } \label{mu(z)m2}
\end{figure}

In Figure \ref{Eosm2}, we can observe the behavior of the DE and total EoS parameter, and the $\Lambda$CDM model. This figure demonstrates the dominance of radiation in the early epoch, followed by a gradual decrease in this dominance and the rise of the DM-dominated era. At the tail end of this sequence, we can observe the present accelerating expansion of the Universe era, with the EoS parameter approaching a value of -1 at late times. Figure \ref{Eosm2} also illustrates the EoS parameter $\omega_{tot}$, which starts at $\frac{1}{3}$ for radiation, decreases to 0 during the matter-dominated period and ultimately reaches -1. Both $\omega_{\Lambda CDM}$ and $\omega_{DE}$ (blue) approach -1 at late times, with the value of $\omega_{DE}$ being -1 at present. Additionally, at present, the density parameters are approximately $\Omega_{m}\approx 0.3$ for DM and $\Omega_{DE}\approx 0.7$ for DE. One thing can be noted the matter-radiation equality observed at $z\approx 3387$ in Figure \ref{densityparametersm2}. Figure \ref{H(z)m2} displays the Hubble rate behaviour as a function of redshift $z$, showing that the model closely resembles the standard $\Lambda$CDM model. In Figure \ref{decelerationparametersm2}, the transition from deceleration to acceleration occurs at $z\approx 0.65$, and the present value of the deceleration parameter is noted as approximately -0.56. Furthermore, Figure \ref{mu(z)m2} presents the modulus function $\mu_{\Lambda CDM}$ for the $\Lambda$CDM model, along with 1048 pantheon data points and the modulus function $\mu(z)$.\\

\begin{figure}[H]
    \centering
\includegraphics[width=78mm]{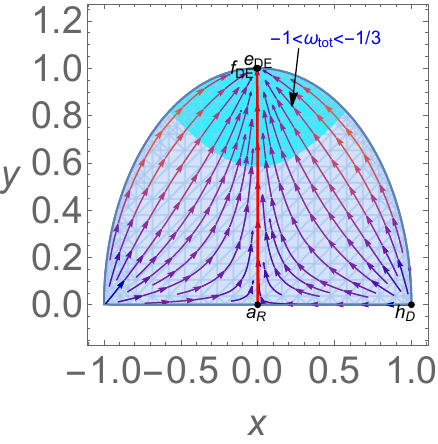}\caption{Visualization of the 2D phase space dynamics with initial conditions and parameter values consistent with Figure \ref{Eosdensitym2}.} \label{2dm2}
\end{figure}

In the 2D phase space portrait represented in Figure \ref{2dm2}, the dynamical variables $x$ and $y$ are plotted. The phase space trajectories demonstrate the behavior of critical points $h_{D}$, $a_{R}$, $d_{M}$, and $b_{R}$ as saddle critical points, while the critical points $e_{DE}$ and $f_{DE}$ exhibit stable behavior. These stable critical points are situated within the region of the accelerating expansion phase of the Universe (quintessence) ($-1<\omega_{tot}<-\frac{1}{3}$), which is highlighted by the cyan color in the upper part of the phase-portrait region.

\subsection{Potential \texorpdfstring{$P_3 \,: V_{0} Sinh^{-\eta}(\beta \phi)$}{}}\label{P_3}

This form of potential plays an important role in analysing the generalise form teleparallel tachyonic model \cite{Bahamonde_2019} and to analyse the value of cosmological constant $\Lambda$ using type Ia Supernovae \cite{SAHNI2000}. Here we consider this potential form to check its physical viability to discuss the evolutionary epochs of the Universe in the generalized teleparallel scalar tensor formalism. The critical points corresponding to this potential has been presented in the following Table \ref{P3criticalpoints}. In this case, the value of the $\alpha_{1}=\frac{1}{\eta}$ and $\alpha_{2}=-\eta \beta^2$, as presented in Table \ref{Potentialfunctions}, hence the autonomous dynamical system will be of five-dimensions.
\begin{table}[H]
    \centering 
    \begin{tabular}{|c |c |c |c| c| c| c|} 
    \hline\hline 
    \parbox[c][0.9cm]{1.3cm}{\textbf{C. P.}
    }& $ \{ x, \, y, \, u, \, \rho, \, \lambda \} $ & $q$ &  \textbf{$\omega_{tot}$}& $\Omega_{r}$& $\Omega_{m}$& $\Omega_{DE}$\\ [0.5ex] 
    \hline\hline 
    \parbox[c][1.3cm]{1.3cm}{$\mathcal{A}_{R}$ } &$\{ 0, 0, 0,  1,  \lambda \}$ & $1$ &  $\frac{1}{3}$& $1$& $0$ & $0$ \\
    \hline
    \parbox[c][1.3cm]{1.3cm}{$\mathcal{B}_{R}$ } & $\left\{\frac{2 \sqrt{\frac{2}{3}}}{\beta  \eta },\frac{2}{\sqrt{3} \beta  \eta },0,\frac{\sqrt{\beta ^2 \eta ^2-4}}{\beta  \eta },\beta  \eta\right\}$ & $1$&  $\frac{1}{3}$& $1-\frac{4}{\beta ^2 \eta ^2}$ & $0$ & $\frac{4}{\beta ^2 \eta ^2}$ \\
    \hline
   \parbox[c][1.3cm]{1.3cm}{$\mathcal{C}_{M}$ } &  $\{0,\, \, 0,\, \, 0, \, \, 0, \, \, \lambda\}$ & $\frac{1}{2}$ &  $0$& $0$ & $1$ & $0$\\
   \hline
   \parbox[c][1.3cm]{1.3cm}{$\mathcal{D}_{M}$} &   $\left\{\frac{\sqrt{\frac{3}{2}}}{\beta  \eta },\frac{\sqrt{\frac{3}{2}}}{\beta  \eta },0,0,\beta  \eta\right\}$ & $\frac{1}{2}$ &  $0$& $0$ & $1-\frac{3}{\beta ^2 \eta ^2}$ & $\frac{3}{\beta ^2 \eta ^2}$\\
   \hline
 \parbox[c][1.3cm]{1.3cm}{$\mathcal{E}_{DE}$} &  $\left\{\frac{\beta  \eta }{\sqrt{6}},\sqrt{1-\frac{\beta ^2 \eta ^2}{6}},0,0,\beta  \eta\right\}$ & $-1+\frac{\beta ^2 \eta ^2}{2}$ & $-1+\frac{\beta ^2 \eta ^2}{3}$ & $0$ & $0$ & $1$\\
 \hline
 \parbox[c][1.3cm]{1.3cm}{$\mathcal{F}_{DE}$} &  $\left\{0,1,\frac{\beta\eta }{3},0,\beta\eta\right\}$ & $-1$ &  $-1$ & $0$ & $0$ & $1$\\
 \hline
 \parbox[c][1.3cm]{1.3cm}{$\mathcal{G}_{DE}$} &  $\left\{ 0,1,0,0,0\right\}$ & $-1$ &  $-1$ & $0$ & $0$ & $1$\\
 \hline
 \parbox[c][1.3cm]{1.3cm}{$\mathcal{H}_{D}$} &  $\left\{1,0,0,0,\beta  \eta\right\}$ & $2$ &  $1$ & $0$ & $0$ & $1$\\
 \hline
    \end{tabular}
    \caption{Critical points analysis: values of $q$,\, $\omega_{tot}$,\, $\Omega_{r}$,\, $\Omega_{m}$ 
 and $\Omega_{DE}$ for potential \ref{P_3} $(P_3)$.}
    \label{P3criticalpoints}
\end{table}
The detailed analysis of the critical points is presented as follows,
\begin{itemize}
\item{\textbf{Radiation-Dominated Critical Points} $\mathcal{A}_R, \mathcal{B}_R$}: \\
The critical point $\mathcal{A}_{R}$ is describing the standard radiation-dominated era with $\Omega_{r}=1$, At this critical point, the value of $q=1, \, \omega_{tot}=\frac{1}{3}$. Point $\mathcal{B}_{R}$ describes the non-standard radiation-dominated era with the small contribution of the DE. It is a scaling solution and the physical viability condition on the parameter $0<\Omega_{DE}<1$ applies the condition on the parameters $\beta, \eta$ as $\Big[\beta <0\land \left(\eta <-2 \sqrt{\frac{1}{\beta ^2}}\lor \eta >2 \sqrt{\frac{1}{\beta ^2}}\right)\Big)$ or $\Big(\beta >0\land \left(\eta <-2 \sqrt{\frac{1}{\beta ^2}}\lor \eta >2 \sqrt{\frac{1}{\beta ^2}}\right)\Big]$. This critical point is appeared points are analysed in \cite{Gonzalez-Espinoza:2020jss}.\\

\item{\textbf{Matter-Dominated Critical Points} $\mathcal{C}_M, \mathcal{D}_M$}: \\
Both of these critical points are in the dark-matter dominated era with $q=\frac{1}{2}, \omega_{tot}=0$. The point $\mathcal{C}_M$ describes the standard matter-dominated era with $\Omega_{m}=1$. The critical point $D_{M}$ represents the scaling matter-dominated solution with $\Omega_{DE}=\frac{3}{\beta^2 \eta^2}$. The physically viable condition applies the condition on model parameters, $\beta, \, \eta$ as $\left[\beta <0\land \left(\eta <-\sqrt{3} \sqrt{\frac{1}{\beta ^2}}\lor \eta >\sqrt{3} \sqrt{\frac{1}{\beta ^2}}\right)\right] $ or $ \left[\beta >0\land \left(\eta <-\sqrt{3} \sqrt{\frac{1}{\beta ^2}}\lor \eta >\sqrt{3} \sqrt{\frac{1}{\beta ^2}}\right)\right]$. This matter-dominated critical point $\mathcal{D}_{M}$ appeared in the study made in the standard quintessence model and the teleparallel DE model \cite{Xu_2012,copelandLiddle}.\\

\item{\textbf{DE-Dominated Critical Points} $\mathcal{E}_{DE}, \mathcal{F}_{DE}, \mathcal{G}_{DE}$}: \\
The critical point $\mathcal{E}_{DE}$ is a DE-dominated critical point with $\Omega_{DE}=1$ and is a late-time scaling solution at which $\omega_{tot}=-1+\frac{\beta^2 \eta^2}{3}$. At this point, $\omega_{tot}$ lies in the quintessence region for\\
$\eta \in \mathbb{R}\land \left[\left(\beta <0\land -\sqrt{2} \sqrt{\frac{1}{\beta ^2}}<\eta <\sqrt{2} \sqrt{\frac{1}{\beta ^2}}\right)\lor \beta =0\lor \left(\beta >0\land -\sqrt{2} \sqrt{\frac{1}{\beta ^2}}<\eta <\sqrt{2} \sqrt{\frac{1}{\beta ^2}}\right)\right]$.\\
Points $\mathcal{F}_{DE}$ and $G_{DE}$ behave like cosmological constant. These are the de-Sitter solutions and represent the standard DE-dominated era of the evolution of the Universe with $\Omega_{DE}=1$.  
 \item{\textbf{Critical Point Representing the Stiff Matter $\mathcal{H}_D$ :}} \\
 The critical point $\mathcal{H}_{D}$ with $\omega_{tot}=1$ describes the stiff matter-dominated era. Although the value of $\Omega_{DE}=1$, this critical point can not describe the DE-dominated era of the evolution of the Universe. This critical point is also analysed in the \cite{Gonzalez_Espinoza_2020}. 

\end{itemize}

\textbf{The Eigenvalues and the Stability Conditions :}
\begin{itemize}
\item \textbf{Stability of Critical Points} $\mathcal{A}_{R}, \mathcal{B}_{R}:$ \\
The eigenvalues at $\mathcal{A}_{R}$ are
$\left[\nu_{1}=2,\, \nu_{2}=-1,\, \nu_{3}=1,\, \nu_{4}=0,\,\nu_{5}=0 \right]$. The presence of the plus and the minus signatures indicates that this critical is a saddle point. The eigenvalues at critical point $\mathcal{B}_{R}$ are
$\Big[\nu_{1}=-\frac{4 \alpha }{\beta  \eta },\, \nu_{2}=-\frac{8}{\eta }, \nu_{3}=1,\, \nu_{4}=-\frac{\sqrt{64 \beta ^4 \eta ^4-15 \beta ^6 \eta ^6}}{2 \beta ^3 \eta ^3}-\frac{1}{2},\,\nu_{5}=\frac{\sqrt{64 \beta ^4 \eta ^4-15 \beta ^6 \eta ^6}}{2 \beta ^3 \eta ^3}-\frac{1}{2} \Big]$ and is saddle at $\left[\alpha <0\land \beta >0\land -\frac{2}{\beta }<\eta <0\right]$.

\item \textbf{Stability of Critical Points} $\mathcal{C}_{M}, \mathcal{D}_{M}:$\\
The eigenvalues at the critical point $\mathcal{C}_{M}$ are $\Big[\nu_{1}=-\frac{3}{2},\,\nu_{2}=\frac{3}{2},\,\nu_{3}=-\frac{1}{2},\,\nu_{4}=0,\,\nu_{5}=0\Big]$, as there is a presence of positive and the negative eigenvalues, this is a saddle critical point. The eigenvalues at the critical point $\mathcal{D}_{M}$ are $\Big[\nu_{1}=-\frac{3 \alpha }{\beta  \eta },\,\nu_{2}=-\frac{6}{\eta },\,\nu_{3}=-\frac{1}{2},\,\nu_{4}=-\frac{3 \sqrt{24 \beta ^4 \eta ^4-7 \beta ^6 \eta ^6}}{4 \beta ^3 \eta ^3}-\frac{3}{4},\,\nu_{5}=\frac{3 \sqrt{24 \beta ^4 \eta ^4-7 \beta ^6 \eta ^6}}{4 \beta ^3 \eta ^3}-\frac{3}{4}\Big]$. This critical point shows stability within the range of the model parameters as $\alpha \in \mathbb{R}\land \alpha \neq 0$ and is showing saddle behaviour for $\alpha <0$.

\item \textbf{Stability of Critical Points} $\mathcal{E}_{DE}, \mathcal{F}_{DE}, \mathcal{G}_{DE}:$\\
The eigenvalues at the critical point $\mathcal{E}_{DE}$ are $\Big[\nu_{1}=-\alpha  \beta  \eta,\,\nu_{2}=-2 \beta ^2 \eta ,\,\nu_{3}=\frac{\beta ^2 \eta ^2}{2}-3,\,\nu_{4}=\frac{\beta ^2 \eta ^2}{2}-2,\,\nu_{5}=\beta ^2 \eta ^2-3\Big]$ 
and it shows stability at $\Big[\alpha <0\land \beta <0\land 0<\eta <\sqrt{3} \sqrt{\frac{1}{\beta ^2}}\Big]\lor \Big[\alpha >0\land \beta >0\land 0<\eta <\sqrt{3} \sqrt{\frac{1}{\beta ^2}}\Big]$,
saddle at $\Big[\alpha <0\land \beta >0\land 0<\eta <\sqrt{3} \sqrt{\frac{1}{\beta ^2}}\Big]\lor \Big[\alpha >0\land \beta <0\land 0<\eta <\sqrt{3} \sqrt{\frac{1}{\beta ^2}}\Big]$ and unstable at $\Big[\alpha <0\land \beta <0\land \eta <-\sqrt{6} \sqrt{\frac{1}{\beta ^2}}\Big]\lor \Big[\alpha >0\land \beta >0\land \eta <-\sqrt{6} \sqrt{\frac{1}{\beta ^2}}\Big]$.\\  The eigenvalues at $\mathcal{F}_{DE}$ are $\Big[\nu_{1}=0, \nu_{2}=-3, \nu_{3}=-2, \nu_{4}=-\frac{3 \sqrt{\left(\beta ^2 \eta ^2+6\right) (\beta  \eta  (8 \alpha +\beta  \eta )+6)}}{2 \left(\beta ^2 \eta ^2+6\right)}-\frac{3}{2},\\ \nu_{5}=\frac{3}{2} \Big[\frac{\sqrt{\left(\beta ^2 \eta ^2+6\right) (\beta  \eta  (8 \alpha +\beta  \eta )+6)}}{\beta ^2 \eta ^2+6}-1\Big]$. The stability at this critical point is analysed using CMT. This critical point shows stable behaviour within the range $\eta \in \mathbb{R}\land ((\lambda<0\land \beta <0)\lor (\lambda>0\land \beta <0))$, and the detailed formalism is presented in the \textbf{Appendix}-\ref{CMT}. At the critical point $\mathcal{G}_{DE}$ have the eigenvalues $\Big[\nu_{1}=-3,\,\nu_{2}=-2,\,\nu_{3}=0,\,\nu_{4}=\frac{1}{2} \left(-\sqrt{12 \beta ^2 \eta +9}-3\right),\,\nu_{5}=\frac{1}{2} \left(\sqrt{12 \beta ^2 \eta +9}-3\right)\Big]$. Based upon the CMT, this critical point shows stable behaviour for  $\alpha>0$, where $\dot{u}$ is negative., and the detailed formalism is presented in \textbf{Appendix}-\ref{CMT}.

 \item \textbf{Stability of Critical Point} $\mathcal{H}_{D}$:\\
 The eigenvalues at critical point $\mathcal{H}_{D}$ are $\Big[\nu_{1}=3,\, \nu_{2}=1,\,\nu_{3}=\sqrt{6} \alpha ,\,\nu_{4}=-2 \sqrt{6} \beta ,\,\nu_{5}=3-\sqrt{\frac{3}{2}} \beta  \eta \Big]$
       This critical point is saddle at $\left[\alpha >0\land \beta >0\land \eta >\frac{\sqrt{6}}{\beta }\right]$ and is unstable at $\left[\alpha <0\land \beta <0\land \eta >\frac{\sqrt{6}}{\beta }\right]$.    
\end{itemize}

\textbf{Numerical Results:}\\

\begin{figure}[H]
\centering
\begin{subfigure}[h]{0.3\textwidth}
\centering
\includegraphics[width=78mm]{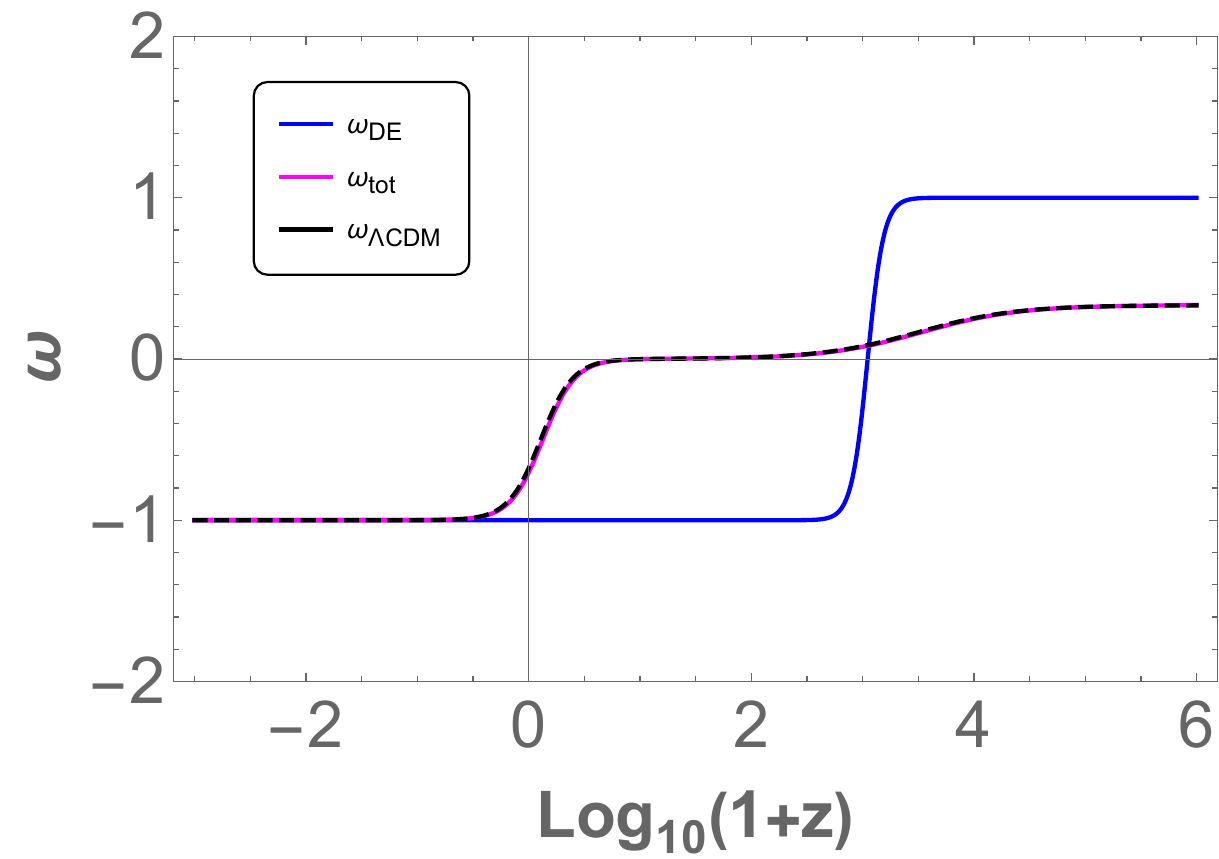}
\caption{The behaviour of EoS parameters.}\label{Eosm3}
\end{subfigure}
\hspace{1.9cm}
\begin{subfigure}[h]{0.5\textwidth}
\centering
\includegraphics[width=78mm]{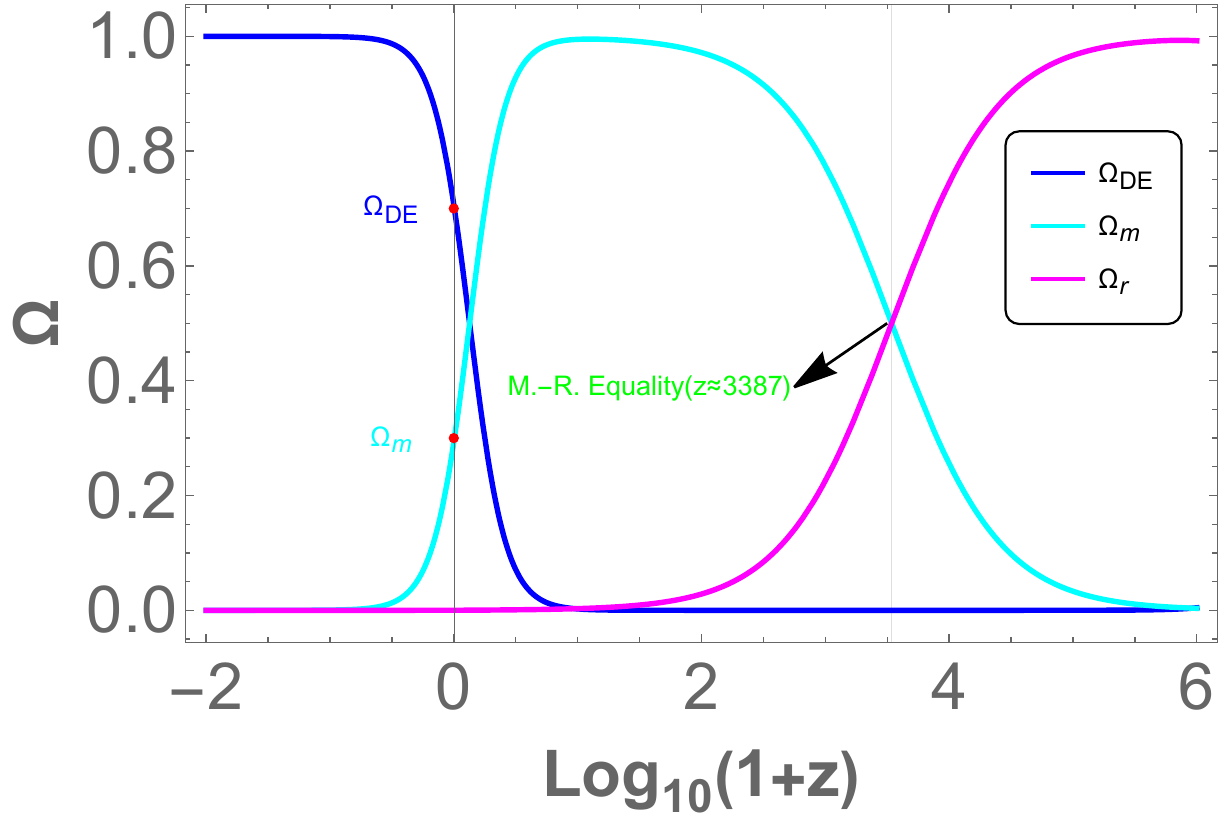}
\caption{The behaviour of standard density parameters.}\label{densityparametersm3}
\end{subfigure}
\caption{The initial conditions are: $x_C=10^{-8.89} ,\,y_C=10^{-2.89} ,\,u_C=10^{-5.96} ,\,\rho_C=10^{-0.75}, \lambda_{c}=10^{-1.3}, \, \alpha=-5.2, \, \eta= -0.2, \, \beta=-0.21$. }\label{Eosdensitym3}
\end{figure}

\begin{figure}[H]
\centering
\begin{subfigure}[h]{0.3\textwidth}
\centering
\includegraphics[width=78mm]{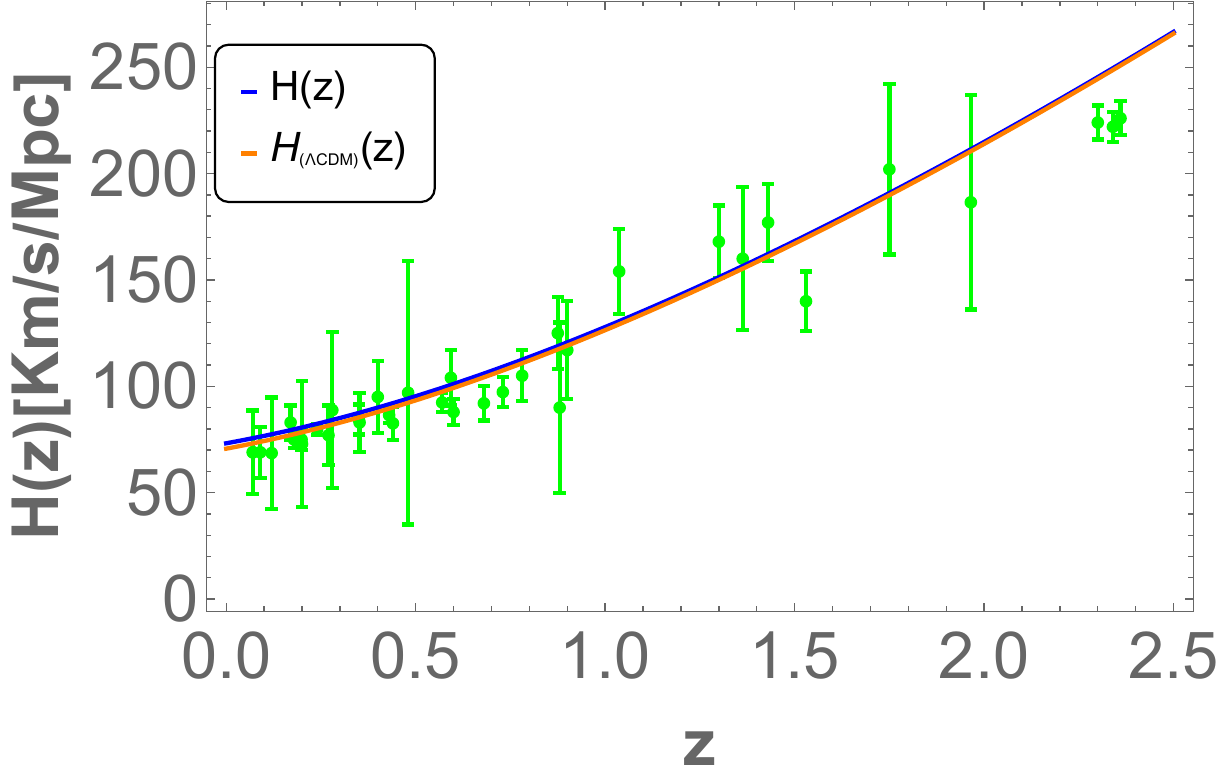}
\caption{The behaviour of Hubble parameter.}\label{H(z)m3}
\end{subfigure}
\hspace{1.9cm}
\begin{subfigure}[h]{0.35\textwidth}
\centering
\includegraphics[width=78mm]{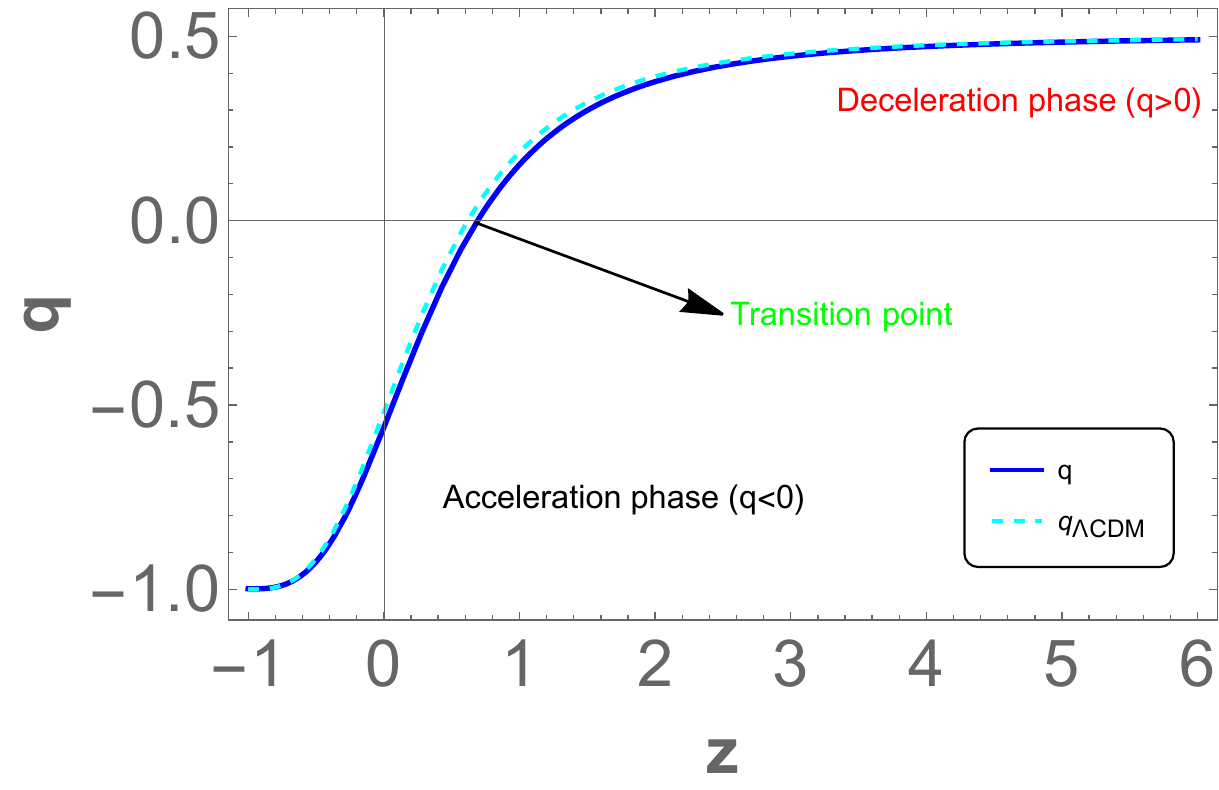}
\caption{The behaviour of deceleration parameter.}\label{decelerationparametersm3}
\end{subfigure}
\caption{The initial conditions are: $x_C=10^{-8.89} ,\,y_C=10^{-2.89} ,\,u_C=10^{-5.96} ,\,\rho_C=10^{-0.75}, \lambda_{c}=10^{-1.3}, \, \alpha=-5.2, \, \eta= -0.2, \, \beta=-0.21$. }\label{h(z)q(z)m3}
\end{figure}

\begin{figure}[H]
    \centering
\includegraphics[width=90mm]{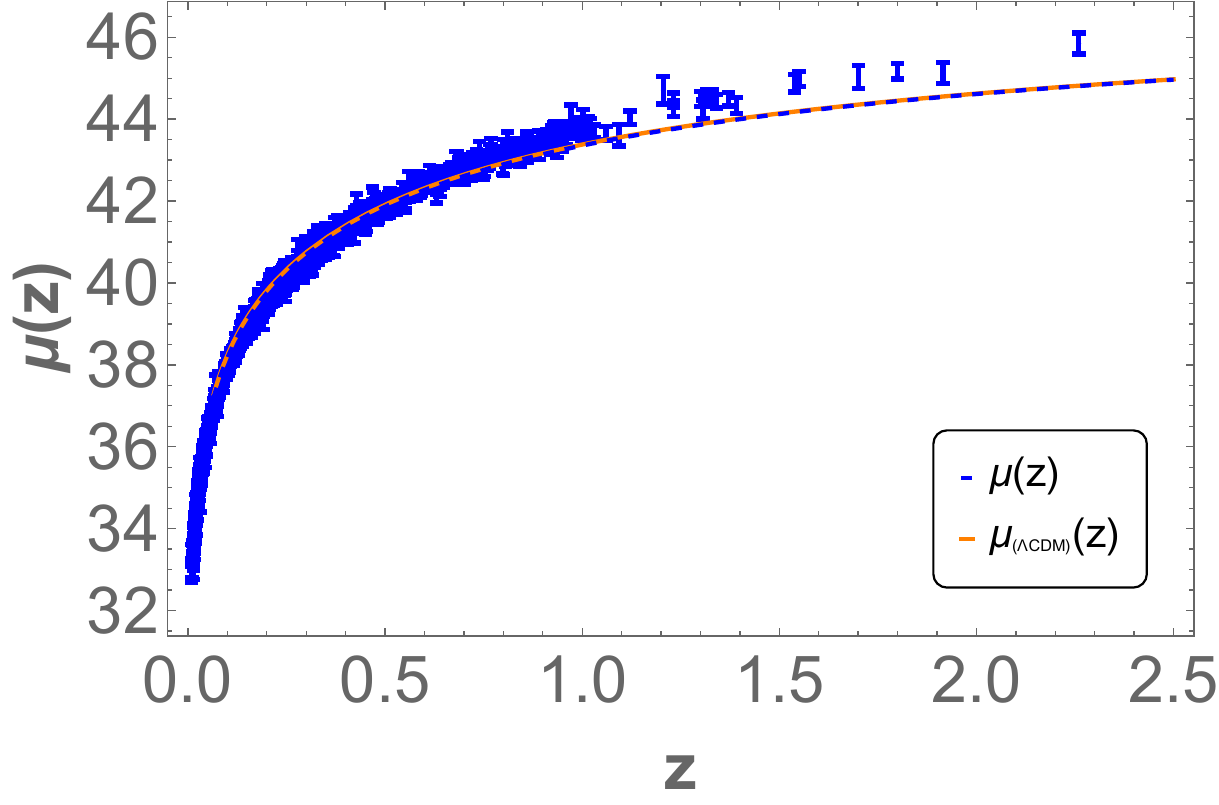}\caption{Plot of the observed distance modulus function $\mu(z)$ and the predicted $\Lambda$CDM model distance modulus function $\mu_{\Lambda CDM}(z)$. The initial conditions are: $x_C=10^{-8.89} ,\,y_C=10^{-2.89} ,\,u_C=10^{-5.96} ,\,\rho_C=10^{-0.75}, \lambda_{c}=10^{-1.3}, \, \alpha=-5.2, \, \eta= -0.2, \, \beta=-0.21$. } \label{mu(z)m3}
\end{figure}

The EoS parameter and standard density parameters are illustrated in Figure \ref{Eosdensitym3}. These plots reveal  the sequence of the dominance of radiation in the early epoch, followed by the dominance of DM, and ultimately the dominance of the DE era at present and in the late stage of the evolution. Currently, the value of the standard DM density parameter $\Omega_{m}\approx 0.3$ and the standard DE density parameter $\Omega_{DE}\approx 0.7$. Their equality is observed at $z\approx 3387$ in Figure \ref{densityparametersm3}. The Figure \ref{Eosm3} depicts the behavior of the EoS parameter $\omega_{tot}$, which starts from $\frac{1}{3}$ for radiation, approaches 0 during the DM-dominated period, and ultimately tends to $-1$. Both $\omega_{\Lambda CDM}$ and $\omega_{DE}$ approach $-1$ at late times, with the current value of $\omega_{DE}$ being equal to $-1$ at $z=0$. The deceleration parameter (q) and $q_{\Lambda CDM}$ depicted in Figure \ref{decelerationparametersm3}, which describes the transition point for deceleration to acceleration, occurring at $z\approx 0.65$. The present value of the deceleration parameter is observed to be $\approx -0.57$. Additionally, Figure \ref{H(z)m3} displays the Hubble rate evolution as a function of redshift $z$, indicating close agreement between the proposed model and the standard $\Lambda$CDM model. Finally, Figure \ref{mu(z)m3} presents the $\Lambda$CDM model modulus function $\mu_{\Lambda CDM}$, 1048 pantheon data points, and the modulus function $\mu(z)$.

\begin{figure}[H]
    \centering
\includegraphics[width=78mm]{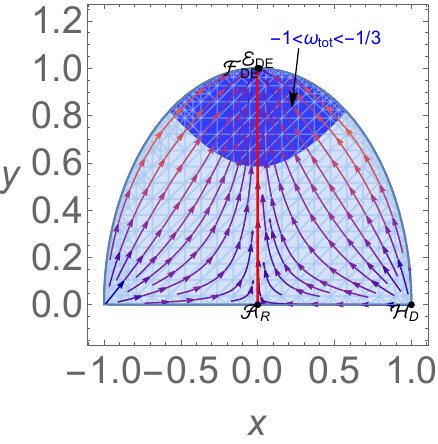}\caption{Visualization of the 2D phase space dynamics with initial conditions and parameter values consistent with Figure \ref{Eosdensitym3}.} \label{2dm3}
\end{figure}

In Figure \ref{2dm3}, we have depicted a 2D phase space diagram illustrating the phase space portrait of the dynamical variables $x$ and $y$. Analysis of trajectories in this plot reveals that the critical points $\mathcal{E}_{DE}$ and $\mathcal{F}_{DE}$ exhibit stable behavior, whereas the critical points $\mathcal{H}_{D}$, $\mathcal{A}_{R}$, $\mathcal{B}_{R}$, and $\mathcal{D}_{M}$ demonstrate saddle behavior. It is noteworthy that the critical points $\mathcal{E}_{DE}$ and $\mathcal{F}_{DE}$ are situated within the region corresponding to the accelerating expansion (quintessence) phase of the Universe, shaded by the blue region in the figure, where ($-1<\omega_{tot}<-\frac{1}{3}$).

\section{Discussion and Conclusion}\label{Conclusion}

A key approach to understanding the background cosmology is to analyze the dynamical systems. Using this technique, we can identify the critical points of a cosmological viable model and their properties. In addition to being validated by observable physics and cosmology, these predictions can also be hinted at through the expanding Universe. Such tests could be performed on a modified gravity model. By linking critical point analysis with stability and phase space images, specific models or parameter ranges within the selected models can be validated or invalidated. In the present work, we have taken into consideration coupling function in the exponential form $g(\phi)=g_{0}e^{-\alpha \phi \kappa}$ to the teleparallel boundary term $B$ and three different forms of the potential function, which is presented in Table \ref{Potentialfunctions}. 
It is also required for the matter and radiation-dominated eras to occur before the DE period for any DE scenario to succeed. The matter and radiation solutions are explained within the context of dynamical systems as critical points of the autonomous system that are
 unstable for radiation,  or saddle for matter. We have discovered novel scaling solutions for the scaling radiation and matter epochs and the crucial points characterizing the standard radiation and matter eras for the DE model under study in this investigation. \\

All three potential functions have stable critical points that explore the accelerated expansion phase of the Universe at a late-time. Scaling solutions have also been obtained for critical points. According to the critical points, non-standard matter and radiation-dominated phases in the Universe have been observed. From our results, we can see that the results match the quintessence model presented in \cite{Xu_2012,copelandLiddle}. We have obtained the matter and DE density parameters at present $z=0$ are found to be $\Omega_{m}^{0}\approx 0.3$ and $\Omega_{de}^{0}\approx 0.7$. Additionally, the matter-radiation equality value is found at $z\approx 3387$. For all three scenarios, the EoS parameter at present and late time have an approach to $-1$ with the present time values aligned with the recent observation \cite{Aghanim:2018eyx}.
Based on the behavior of Figure \ref{decelerationparametersm1}, we can claim that for the exponential form of $V(\phi)$, the deceleration parameter displays the transition from deceleration to acceleration phase at $z=0.66$. Its current value is $q(z=0)=-0.53$. We obtained the transition point at $z=0.65$ for the $V(\phi)=Cosh(\xi \phi)-1$, and the current value of the deceleration parameter is $q(z=0)=-0.56$ [Figure \ref{decelerationparametersm2}]. We find the transition point at $z=0.61$ for the $V(\phi)=V_{0}Sinh^{-\eta}(\beta \phi)$, and the current value of the deceleration parameter is $q(z=0)=-0.57$ can be observed from Figure \ref{decelerationparametersm3}. In each of the three scenarios, the deceleration parameter value and transition point matched cosmological findings \cite{PhysRevD.90.044016a, PhysRevResearch.2.013028}. We compared our results in all three potential functions with the Hubble 31 data points \cite{Moresco_2022_25} and the Supernovae Ia data 1048 data points \cite{Scolnic_2018}. We may conclude that the outcomes of our model closely resemble those of the conventional $\Lambda$CDM model based on the behavior of the Figures [\ref{H(z)m1}, \ref{H(z)m2}, \ref{H(z)m3}]. The modulus function of our models was shown alongside the 1048 Supernovae Ia data points in Figure [\ref{mu(z)m1}, \ref{mu(z)m2}, \ref{mu(z)m3}], using the conventional $\Lambda$CDM model modulus function. The outcomes closely align with the $\Lambda$CDM model. The quintessence region is shaded and is pointed using an arrow in the 2-d phase space diagrams presented in Figures \ref{2dm1}, \ref{2dm2}, and 
\ref{2dm3}. The phase space trajectories move from the early-time decelerating phase to the stable late-time DE solutions. The present value of different standard density parameters, EoS parameters, and $q$ for all three cases are presented in Table \ref{Potentialfunctions}.\\

\begin{table}[H]
    \centering 
    \begin{tabular}{|c |c |c |c| c| c| c| c|} 
    \hline 
 \multicolumn{8}{|c|}{\textbf{Results of the potential functions}} \\
    \hline 
    \parbox[c][0.9cm]{0.9cm}  &\textbf{Potential function} $V(\phi)$ & $\Omega_{DE}^0 \approx$ & $\Omega_{m}^0 \approx$&  \begin{tabular}{@{}c@{}}matter-radiation \\equality $(z_{eq}) \approx$   \end{tabular}&$\omega_{DE}^0\approx$ & $q^{0}$& \begin{tabular}{@{}c@{}}transition point \\$(z_{transi.}) \approx$ \end{tabular} \\ [0.5ex] 
    \hline 
    \parbox[c][0.9cm]{0.9cm} {$P_1$} &$V_{0}e^{-\kappa \phi}$ & $0.7$&  $0.3$& $3387$ & $-1$ & $-0.53$&0.66 \\
    \hline
   \parbox[c][0.9cm]{0.9cm} {$P_2$} & $Cosh(\xi \phi)-1$ & $0.7$ &  $0.3$& $3387$ & $-1$&$-0.56$&0.65 \\
   \hline
   \parbox[c][0.9cm]{0.9cm }{$P_3$} & $V_{0} Sinh^{-\eta}(\beta \phi)$ & $0.7$ &  $0.3$& $3387$ & $-1$&$-0.57$&0.61 \\
   \hline
    \end{tabular}
    \caption{Result summary, in the table upper indices $0$ define present time at $z=0$.}
    \label{Potentialfunctions}
\end{table}

 This work demonstrates that the class of potentials may describe the accelerating expansion of the Universe. Thus, the selection of potentials is still arbitrary. Even though three distinct potentials were chosen for this research, one may think of more potentials to perform the same. To keep the analysis from getting too lengthy, we have limited it to just three cases. Despite this, we have determined the stability regions of the parameters of each potential function form. All three potential functions show early-time deceleration and late-time cosmic acceleration through the behavior of the critical points, which indicates that the late-time cosmic acceleration-based critical points are stable. Moreover, we have observed that the exponential form of potential form does not have roots of $f=\lambda (\Gamma-1)$, which means the autonomous system is reduced to four dimensions. However, two other potential function roots are not zero, so in these cases, $\frac{d\lambda}{dN} \neq 0$ indicates a five-dimensional autonomous system.

This work shows promise; more work in the cosmic setting, utilizing either perturbation theory or observational
constraint analysis, should be done to explore further. This might shed further light on this idea and its connections
to the Universe’s large-scale structure and the cosmic microwave background radiation power spectrum.

\section{Appendix}\label{Appendix}
\subsection{Datasets}\label{Datasets}
\subsubsection{\bf{Hubble data $H(z)$}}

In this work, we have taken 31 data points \cite{Moresco_2022_25} to describe the behavior of the Hubble rate in our model. The standard $\Lambda$CDM model will also be compared to our model. We know,
 
\begin{equation}\label{hubble_LCDM}
H_{\Lambda CDM}= H_{0}\sqrt{(1+z)^3 \Omega_{m}+(1+z)^4 \Omega_{r}+\Omega_{de}} \,,   
\end{equation}

\subsubsection*{\bf{Supernovae Ia}}
Another component of our baseline data set is the Pantheon compilation of 1048 SNIa distance measurements spanning $0.01<z<2.3$ redshifts \cite{Scolnic_2018}. This dataset incorporates observations from prominent programs such as PanSTARRS1, Hubble Space Telescope (HST) survey, SNLS, and SDSS. By amalgamating data from diverse sources, the Pantheon collection offers valuable insights into the properties and behaviors of Type Ia supernovae and their cosmic implications. Furthermore, it demonstrates the use of stellar luminosity as a means of determining distances in an expanding Universe, with the distance moduli function being a key component of this analysis, and is represented as,
\begin{equation}\label{panmoduli}
\mu(z_{i}, \Theta)=5 \log_{10}[D_{L}(z_i, \Theta)]+M    
\end{equation}
Here, $M$ is the nuisance parameter, while $D_{L}$ denotes the luminosity distance. Luminosity distance can be calculated as follows: 
\begin{equation}\label{luminositydistance}
D(z_{i}, \Theta)=c (1+z_{i}) \int^{z_i}_{0} \frac{dz}{H(z, \Theta)} \,,   
\end{equation}

\subsection{Center Manifold Theory (CMT) for non-hyperbolic critical points}\label{CMT}
Central manifold theory (CMT) is a branch of dynamical systems theory that deals with the behavior of systems near fixed points. The mathematical framework of CMT was explained by Perko \cite{Perko2001}. The linear stability theory fails to explain the stability of critical points if their eigenvalues include zero eigenvalues. An analysis of stability is possible in a CMT because the dimensionality of the system reduced near that point. When the system passes through the critical point, it behaves in an invariant local center manifold $W^c$. The central manifold $W^c$ is an invariant manifold associated with eigenvalues having zero real parts. The dynamics on this manifold capture the essential features of the behavior of system near the equilibrium.

Let \( f \in C^{r}(E) \), with \( E \) being an open subset of \( \mathbb{R}^{n} \) that includes the origin, and \( r \geq 1 \). Suppose \( f(0) = 0 \) and the derivative \( Df(0) \) has \( c \) eigenvalues with zero real parts and \( s \) eigenvalues with negative real parts, where \( c + s = n \). Typically, the system can be rewritten in the following form.

\begin{eqnarray}
\dot{x}= \mathcal{A}x + \mathcal{F}(x,y) \nonumber \\ 
\dot{y}=  \mathcal{B}y + \mathcal{G}(x,y) \label{CMT1}    
\end{eqnarray}

The ordered pair $(x, y) \in R^c \times R^s$ is considered, where $\mathcal{A}$ is a square matrix with $c$ eigenvalues having zero real parts, $\mathcal{B}$ is a square matrix with $s$ eigenvalues having negative real parts; and $\mathcal{F}(0) = \mathcal{A}(0) = 0, D\mathcal{A} (0) = D\mathcal{G}(0) = 0$. Additionally, there exists a small positive value $\epsilon>0$ and a function $g(x)$ that belongs to $C^r (N_{\epsilon}$(0)), defining the local center manifold and meeting specific conditions:

\begin{eqnarray}
[\mathcal{A}x+\mathcal{F}(x,g(x))] 
 -\mathcal{B}h(x)-\mathcal{G}(x,g(x))=Dg(x)=\mathcal{N}(g(x))=0 \label{CMT2}    
\end{eqnarray}

for $|x|< \epsilon$; and the center manifold can be derived using the system of differential equations as follows,

\begin{eqnarray}
 \dot{x}= \mathcal{A}x + \mathcal{F}(x,g(x)) \label{CMT3}   
\end{eqnarray}

for all $x\in \mathbb{R}^{c}$ with $|x|<\epsilon$.\\

\textbf{CMT for critical point $G_{DE}$:}\\

We have obtained the Jacobian matrix for the critical point \( G_{DE} \) for the autonomous system presented in Eq. (\ref{dynamicalsystem}) as follows:

\[
J(G_{DE}) = 
\begin{bmatrix}
 -3 & 0  &-3\sqrt{\frac{3}{2}}  & 0  \\
 0 & -3 & 0 & 0  \\
 0 & 0 & 0 & 0  \\
 0 & 0 & 0 & -2 
\end{bmatrix}
\]

The eigenvalues of the above Jacobian matrix \( G_{DE} \) are \( \nu_{1} = -2 \), \( \nu_{2} = 0 \), \( \nu_{3} = -3 \), and \( \nu_{4} = -3 \) as obtained in Sec. \ref{P_1}. The eigenvectors corresponding to these eigenvalues are

\begin{math}
\left[0,1,0,0\right]^T
\end{math} \begin{math}
\left[1,0,0,0\right]^T
\end{math} \begin{math}
,\left[0,0,0,1\right]^T
\end{math} \begin{math}
,\left[-\sqrt{\frac{3}{2}},0,1,0\right]^T
\end{math}

Now, applying the CMT, we analyze the stability of this critical point \( G_{DE} \). We use the transformation \( X = x \), \( Y = y -1 \), \( Z = u \), and \( R = \rho \) to shift this critical point to the origin. The resultant equations obtained in the new coordinate system can then be written as,

\begin{align}
\begin{pmatrix}
\dot{X}\\ 
\dot{Y} \\ 
 \dot{R}\\ 
 \dot{Z} 
\end{pmatrix}= 
\begin{pmatrix}
-3 & 0 & 0 & 0 \\
0 & -3 & 0 & 0 \\
0 & 0 & -2 & 0  \\
0 & 0 & 0 & 0 \\
\end{pmatrix} 
\begin{pmatrix}
X\\ 
 Y\\ 
 R\\ 
 Z   
\end{pmatrix}+\begin{pmatrix}
 non\\ linear\\ term   
\end{pmatrix} 
\end{align}

Upon comparing the diagonal matrix with the standard form \eqref{CMT1}, it is apparent that the variables \( X \), \( Y \), and \( R \) demonstrate stability, while \( Z \) serves as the central variable. At this critical point, the matrices \( \mathcal{A} \) and \( \mathcal{B} \) adopt the following form,

\[
\mathcal{A} =
\begin{bmatrix}
 -3 & 0  &0    \\
  0 & -3  &0   \\
 0 & 0 & -2  \\
 
\end{bmatrix}
\hspace{0.5cm}
\mathcal{B} = 
\begin{bmatrix}
0    
\end{bmatrix}
\]

As per the CMT, the manifold can be expressed using a function that is continuously differentiable. We make the assumption that the stable variables can be represented by the following functions: \( X = g_{1}(Z) \), \( Y = g_{2}(Z) \), and \( R = g_{3}(Z) \). By using equation \eqref{CMT2}, we can derive the zeroth-order approximation of the manifold functions in the following manner:
\begin{eqnarray}
\mathcal{N}(g_1(Z)) = \frac{3 \sqrt{6} Z}{3 Z^2+2}\,, \hspace{0.5cm} \mathcal{N}(g_2(Z)) =-\frac{9 Z^2}{3 Z^2+2}\,, \hspace{0.5cm} \mathcal{N}(g_3(Z)) =0 \,.
\end{eqnarray}
In this case, the center manifold is given by the following expression:
\begin{equation}\label{CMTmanifold1}
\dot{Z}= -\frac{18 \alpha  Z^2}{3 Z^2+2}+ higher \hspace{0.15cm} order \hspace{0.15cm} term   
\end{equation}
By CMT, the critical point $G_{DE}$ exhibits stable behavior for \( Z \neq 0 \) and \( \alpha > 0 \), where \( \dot{Z} \) is negative.
\\

\textbf{CMT for critical point $f_{DE}$:}\\

The Jacobian matrix at the critical point \( f_{DE} \) for the autonomous system (\ref{dynamicalsystem}) is given as,\\
$J(f_{DE}) = 
\left(
\begin{array}{ccccc}
 -\frac{12}{\frac{2 \xi ^2}{3}+4} & \frac{2 \sqrt{6} \xi }{\frac{2 \xi ^2}{3}+4} & -\frac{6 \sqrt{6}}{\frac{2 \xi ^2}{3}+4} & 0 & \frac{2 \sqrt{6}}{\frac{2 \xi ^2}{3}+4} \\
 \frac{\sqrt{6} \xi }{\frac{\xi ^2}{3}+2}-\sqrt{\frac{3}{2}} \xi  & -\frac{2 \xi ^2}{\frac{\xi ^2}{3}+2}-\frac{6}{\frac{\xi ^2}{3}+2} & \frac{3 \xi }{\frac{\xi ^2}{3}+2} & 0 & -\frac{\xi }{\frac{\xi ^2}{3}+2} \\
 -\sqrt{\frac{2}{3}} \alpha  \xi  & 0 & 0 & 0 & 0 \\
 0 & 0 & 0 & -\frac{2 \xi ^2}{3 \left(\frac{\xi ^2}{3}+2\right)}-\frac{4}{\frac{\xi ^2}{3}+2} & 0 \\
 0 & 0 & 0 & 0 & 0 \\
\end{array}
\right)$

The eigenvalues of Jacobian matrix $f_{DE}$ are $\nu_{1}=0,\, \nu_{2}=-3,\, \nu_{3}=-2,\, \nu_{4}=-\frac{3 \left(\sqrt{\left(\xi ^2+6\right) \left(8 \alpha  \xi +\xi ^2+6\right)}+\xi ^2+6\right)}{2 \left(\xi ^2+6\right)},\, \nu_{5}= \frac{3}{2} \left(\frac{\sqrt{\left(\xi ^2+6\right) \left(8 \alpha  \xi +\xi ^2+6\right)}}{\xi ^2+6}-1\right)$.\\

The corresponding eigenvectors are 

\begin{math}
\left[0,0,\frac{1}{3},0,1\right]^T
\end{math} \begin{math}
,\left[\frac{3 \sqrt{\frac{3}{2}}}{\alpha  \xi },\frac{3 (2 \alpha -\xi )}{2 \alpha  \xi },1,0,0\right]^T
\end{math} \begin{math}
,\left[0,0,0,1,0\right]^T,
\\
\left[\frac{3 \sqrt{\frac{3}{2}} \left(-\sqrt{\left(\xi ^2+6\right) \left(8 \alpha  \xi +\xi ^2+6\right)}+\xi ^2+6\right)}{2 \alpha  \xi  \left(\xi ^2+6\right)},\frac{3 \xi  \left(4 \alpha  \xi ^2+\xi  \sqrt{\left(\xi ^2+6\right) \left(8 \alpha  \xi +\xi ^2+6\right)}+24 \alpha -\xi ^3-6 \xi \right)}{2 \alpha  \left(\xi ^2+6\right) \left(\sqrt{\left(\xi ^2+6\right) \left(8 \alpha  \xi +\xi ^2+6\right)}+3 \xi ^2+6\right)},1,0,0\right]^T
\end{math}
and, \\ 
\begin{math}
\left[ \frac{3 \sqrt{\frac{3}{2}} \left(\sqrt{\left(\xi ^2+6\right) \left(8 \alpha  \xi +\xi ^2+6\right)}+\xi ^2+6\right)}{2 \alpha  \xi  \left(\xi ^2+6\right)},-\frac{3 \xi  \left(4 \alpha  \xi ^2-\xi  \sqrt{\left(\xi ^2+6\right) \left(8 \alpha  \xi +\xi ^2+6\right)}+24 \alpha -\xi ^3-6 \xi \right)}{2 \alpha  \left(\xi ^2+6\right) \left(\sqrt{\left(\xi ^2+6\right) \left(8 \alpha  \xi +\xi ^2+6\right)}-3 \xi ^2-6\right)},1,0,0 \right]^T  
\end{math}\\

To shift the critical points to the origin, the specific transformations we employed here are: \( X = x \), \( Y =  y -1 \), \( Z = u - \frac{\xi}{3} \), \( R = \rho \), and \( L = \lambda - \xi \). In this new coordinate system, we expressed the equations in the following form:

\begin{align}
\begin{pmatrix}
\dot{X}\\ 
\dot{Y} \\ 
 \dot{Z}\\ 
 \dot{R} \\
 \dot{L}
\end{pmatrix}= 
\left(
\begin{array}{ccccc}
-\frac{3 \left(\sqrt{\left(\xi ^2+6\right) \left(8 \alpha  \xi +\xi ^2+6\right)}+\xi ^2+6\right)}{2 \left(\xi ^2+6\right)} &0& 0& 0 &0 \\
 0  &  \frac{3}{2} \left(\frac{\sqrt{\left(\xi ^2+6\right) \left(8 \alpha  \xi +\xi ^2+6\right)}}{\xi ^2+6}-1\right)& 0 & 0 &0\\
0 & 0 & -3 & 0 & 0 \\
 0 & 0 & 0 & -2 & 0 \\
 0 & 0 & 0 & 0 & 0 \\
\end{array}
\right)
\begin{pmatrix}
X\\ 
 Y\\ 
 Z\\ 
 R\\ 
 L
\end{pmatrix}+\begin{pmatrix}
 non\\ linear\\ term   
\end{pmatrix} 
\end{align}

We can determine that \( X \), \( Y \), \( Z \), and \( R \) are the stable variables, while \( L \) is the central variable. At this critical point, the matrices \( \mathcal{A} \) and \( \mathcal{B} \) take the following form:

\[
\mathcal{A} =\left(
\begin{array}{ccccc}
-\frac{3 \left(\sqrt{\left(\xi ^2+6\right) \left(8 \alpha  \xi +\xi ^2+6\right)}+\xi ^2+6\right)}{2 \left(\xi ^2+6\right)} &0& 0& 0  \\
 0  &  \frac{3}{2} \left(\frac{\sqrt{\left(\xi ^2+6\right) \left(8 \alpha  \xi +\xi ^2+6\right)}}{\xi ^2+6}-1\right)& 0 & 0 \\
0 & 0 & -3 & 0  \\
 0 & 0 & 0 & -2  \\
\end{array}
\right)
\hspace{0.5cm}
\mathcal{B} = 
\begin{bmatrix}
0    
\end{bmatrix}
\]

According to CMT, we have made specific assumptions that  $X=g_{1}(L)$, $Y=g_{2}(L)$, $Z=g_{3}(L)$, and $R=g_{4}(L)$ are the stable variables. Now By using equation \eqref{CMT2}, we have derived the zeroth order approximation of the manifold functions.

\begin{eqnarray}
\mathcal{N}(g_1(L)) = -\frac{3 \sqrt{6} L}{\xi ^2+6}\,, \hspace{0.5cm} \mathcal{N}(g_2(L)) =\frac{3 L \xi }{\xi ^2+6}\,, \hspace{0.5cm} \mathcal{N}(g_3(L)) =0 \,,\hspace{0.5cm} \mathcal{N}(g_4(L)) =0 \,
\end{eqnarray}

With these, the central manifold can be obtained as 
\begin{equation}\label{CMTmanifold2}
\dot{L}= -\frac{18 L^2 \xi }{\xi ^2+6}+ higher \hspace{0.2cm} order \hspace{0.2cm} term   
\end{equation}

By applying the CMT, this critical point shows stable behavior for  $ \xi >0$, where $\dot{L}$ is negative.\\

\textbf{CMT for critical point $g_{DE}$:}\\

The Jacobian matrix at the critical point $g_{DE}$ for the autonomous system (\ref{dynamicalsystem} ) is as follows:

$J(g_{DE}) = 
\left(
\begin{array}{ccccc}
 -3 & 0 & -3 \sqrt{\frac{3}{2}} & 0 & \sqrt{\frac{3}{2}} \\
 0 & -3 & 0 & 0 & 0 \\
 0 & 0 & 0 & 0 & 0 \\
 0 & 0 & 0 & -2 & 0 \\
 -\sqrt{\frac{3}{2}} \xi ^2 & 0 & 0 & 0 & 0 \\
\end{array}
\right)$

The eigenvalues at critical point $g_{DE}$ are $\nu_{1}=-3,\, \nu_{2}=-2,\, \nu_{3}=0,\, \nu_{4}=\frac{1}{2} \left(-\sqrt{9-6 \xi ^2}-3\right),\, \nu_{5}=\frac{1}{2} \left(\sqrt{9-6 \xi ^2}-3\right) $. The eigenvectors corresponding to these eigenvalues are 
\begin{math}
   \left[0,1,0,0,0\right]^{T} 
\end{math},\\
\begin{math}
   \left[0,0,0,1,0\right]^{T} 
\end{math}
\begin{math}
   \left[0,0,\frac{1}{3},0,1\right]^{T} 
\end{math}
\begin{math}
   \left[-\frac{-\sqrt{2} \sqrt{3-2 \xi ^2}-\sqrt{6}}{2 \xi ^2},0,0,0,1\right]^{T} 
\end{math}
\begin{math}
\left[ -\frac{\sqrt{2} \sqrt{3-2 \xi ^2}-\sqrt{6}}{2 \xi ^2},0,0,0,1\right]^{T} 
\end{math}

To apply CMT, we have shifted the critical point to the origin using a shifting transformations $X=x$, $Y=y-1$, $Z=u$, $R=\rho$, and $L=\lambda$. Then we can write equations in the new coordinate system as, 

\begin{align}
\begin{pmatrix}
\dot{X}\\ 
\dot{Y} \\ 
 \dot{R}\\ 
 \dot{L} \\
 \dot{Z}
\end{pmatrix}= 
\left(
\begin{array}{ccccc}
\frac{1}{2} \left(-\sqrt{9-6 \xi ^2}-3\right) &0& 0& 0 &0 \\
 0  &  \frac{1}{2} \left(\sqrt{9-6 \xi ^2}-3\right)& 0 & 0 &0\\
0 & 0 & -3 & 0 & 0 \\
 0 & 0 & 0 & 0-2 & 0 \\
 0 & 0 & 0 & 0 & 0 \\
\end{array}
\right)
\begin{pmatrix}
X\\ 
 Y\\ 
 R\\ 
 L\\ 
 Z
\end{pmatrix}+\begin{pmatrix}
 non\\ linear\\ term   
\end{pmatrix} 
\end{align}

The variables $X, Y, R$, and $L$ are stable, whereas $Z$ is the central variable. At this critical point, the matrices $\mathcal{A}$ and $\mathcal{B}$ take on the following form:

\[
\mathcal{A} =\left(
\begin{array}{ccccc}
\frac{1}{2} \left(-\sqrt{9-6 \xi ^2}-3\right) &0& 0& 0  \\
 0  &  \frac{1}{2} \left(\sqrt{9-6 \xi ^2}-3\right)& 0 & 0 \\
0 & 0 & -3 & 0  \\
 0 & 0 & 0 & 0-2  \\
\end{array}
\right)
\hspace{0.5cm}
\mathcal{B} = 
\begin{bmatrix}
0    
\end{bmatrix}
\]

Using CMT, we assume the : \( X = g_{1}(Z) \), \( Y = g_{2}(Z) \), \( R = g_{3}(Z) \), and \( L = g_{4}(Z) \) are the stable variables, and the zeroth approximation of the manifold functions are as follows:

\begin{eqnarray}
\mathcal{N}(g_1(Z)) = \frac{3 \sqrt{6} Z}{3 Z^2+2}\,, \hspace{0.5cm} \mathcal{N}(g_2(Z)) =-\frac{9 Z^2}{3 Z^2+2}\,, \hspace{0.5cm} \mathcal{N}(g_3(Z)) =0 \,,\hspace{0.5cm} \mathcal{N}(g_4(Z)) =0 \,
\end{eqnarray}

With these assumptions, the central manifold can be obtained as follows:
 
\begin{equation}\label{CMTmanifold3}
\dot{Z}=-\frac{18 \alpha  Z^2}{3 Z^2+2}+ higher \hspace{0.2cm} order \hspace{0.2cm} term   
\end{equation}

According to CMT, this critical point demonstrates stable behavior for \( Z \ne 0 \) and \( \alpha > 0 \), where \( \dot{Z} \) is negative.
\\

\textbf{CMT for critical point $\mathcal{F}_{DE}$:}\\

The Jacobian matrix at the critical point $\mathcal{F}_{DE}$ for the autonomous system (\ref{dynamicalsystem} ) is,

$J(\mathcal{F}_{DE}) = 
\left(
\begin{array}{ccccc}
 -\frac{12}{\frac{2 \beta ^2 \eta ^2}{3}+4} & \frac{2 \sqrt{6} \beta  \eta }{\frac{2 \beta ^2 \eta ^2}{3}+4} & -\frac{6 \sqrt{6}}{\frac{2 \beta ^2 \eta ^2}{3}+4} & 0 & \frac{2 \sqrt{6}}{\frac{2 \beta ^2 \eta ^2}{3}+4} \\
 \frac{\sqrt{6} \beta  \eta }{\frac{\beta ^2 \eta ^2}{3}+2}-\sqrt{\frac{3}{2}} \beta  \eta  & -\frac{2 \beta ^2 \eta ^2}{\frac{\beta ^2 \eta ^2}{3}+2}-\frac{6}{\frac{\beta ^2 \eta ^2}{3}+2} & \frac{3 \beta  \eta }{\frac{\beta ^2 \eta ^2}{3}+2} & 0 & -\frac{\beta  \eta }{\frac{\beta ^2 \eta ^2}{3}+2} \\
 -\sqrt{\frac{2}{3}} \alpha  \beta  \eta  & 0 & 0 & 0 & 0 \\
 0 & 0 & 0 & -\frac{2 \beta ^2 \eta ^2}{3 \left(\frac{\beta ^2 \eta ^2}{3}+2\right)}-\frac{4}{\frac{\beta ^2 \eta ^2}{3}+2} & 0 \\
 0 & 0 & 0 & 0 & 0 \\
\end{array}
\right)$

The eigenvalues at $\mathcal{F}_{DE}$ are $\nu_{1}=0, \nu_{2}=-3, \nu_{3}=-2, \nu_{4}=-\frac{3 \sqrt{\left(\beta ^2 \eta ^2+6\right) (\beta  \eta  (8 \alpha +\beta  \eta )+6)}}{2 \left(\beta ^2 \eta ^2+6\right)}-\frac{3}{2},\\ \nu_{5}=\frac{3}{2} \Big[\frac{\sqrt{\left(\beta ^2 \eta ^2+6\right) (\beta  \eta  (8 \alpha +\beta  \eta )+6)}}{\beta ^2 \eta ^2+6}-1$.\\
The eigenvectors corresponding to these eigenvalues are \\
\begin{math}
   \left[0,0,\frac{1}{3},0,1\right]^{T} 
\end{math},
\begin{math}
   \left[\frac{3 \sqrt{\frac{3}{2}}}{\alpha  \beta  \eta },\frac{3 (2 \alpha -\beta  \eta )}{2 \alpha  \beta  \eta },1,0,0\right]^{T} 
\end{math}
\begin{math}
   \left[0,0,0,1,0\right]^{T} 
\end{math}\\
\begin{math}
   \left[\frac{3 \sqrt{\frac{3}{2}} \left(-\sqrt{\left(\beta ^2 \eta ^2+6\right) \left(8 \alpha  \beta  \eta +\beta ^2 \eta ^2+6\right)}+\beta ^2 \eta ^2+6\right)}{2 \alpha  \beta  \eta  \left(\beta ^2 \eta ^2+6\right)},\frac{3 \beta  \eta  \left(4 \alpha  \beta ^2 \eta ^2+\beta  \eta  \sqrt{\left(\beta ^2 \eta ^2+6\right) \left(8 \alpha  \beta  \eta +\beta ^2 \eta ^2+6\right)}+24 \alpha -\beta ^3 \eta ^3-6 \beta  \eta \right)}{2 \alpha  \left(\beta ^2 \eta ^2+6\right) \left(\sqrt{\left(\beta ^2 \eta ^2+6\right) \left(8 \alpha  \beta  \eta +\beta ^2 \eta ^2+6\right)}+3 \beta ^2 \eta ^2+6\right)},1,0,0\right]^{T} 
\end{math}\\
\begin{math}
\left[\frac{3 \sqrt{\frac{3}{2}} \left(\sqrt{\left(\beta ^2 \eta ^2+6\right) \left(8 \alpha  \beta  \eta +\beta ^2 \eta ^2+6\right)}+\beta ^2 \eta ^2+6\right)}{2 \alpha  \beta  \eta  \left(\beta ^2 \eta ^2+6\right)},-\frac{3 \beta  \eta  \left(4 \alpha  \beta ^2 \eta ^2-\beta  \eta  \sqrt{\left(\beta ^2 \eta ^2+6\right) \left(8 \alpha  \beta  \eta +\beta ^2 \eta ^2+6\right)}+24 \alpha -\beta ^3 \eta ^3-6 \beta  \eta \right)}{2 \alpha  \left(\beta ^2 \eta ^2+6\right) \left(\sqrt{\left(\beta ^2 \eta ^2+6\right) \left(8 \alpha  \beta  \eta +\beta ^2 \eta ^2+6\right)}-3 \beta ^2 \eta ^2-6\right)},1,0,0\right]^{T} 
\end{math}\\

The transformations used in this case are: \( X = x \), \( Y = y -1 \), \( Z = u - \frac{\beta \eta}{3} \), \( R = \rho \), and \( L = \lambda - \beta \eta \), and in the new coordinate system, the equations can be written as,

\begin{align}
\begin{pmatrix}
\dot{X}\\ 
\dot{Y} \\ 
 \dot{Z}\\ 
 \dot{R} \\
 \dot{L}
\end{pmatrix}= 
\left(
\begin{array}{ccccc}
-\frac{3 \sqrt{\left(\beta ^2 \eta ^2+6\right) (\beta  \eta  (8 \alpha +\beta  \eta )+6)}}{2 \left(\beta ^2 \eta ^2+6\right)}-\frac{3}{2} &0& 0& 0 &0 \\
 0  &  \frac{3}{2} \Big[\frac{\sqrt{\left(\beta ^2 \eta ^2+6\right) (\beta  \eta  (8 \alpha +\beta  \eta )+6)}}{\beta ^2 \eta ^2+6}-1\Big]& 0 & 0 &0\\
0 & 0 & -3 & 0 & 0 \\
 0 & 0 & 0 & 0-2 & 0 \\
 0 & 0 & 0 & 0 & 0 \\
\end{array}
\right)
\begin{pmatrix}
X\\ 
 Y\\ 
 Z\\ 
 R\\ 
 L
\end{pmatrix}+\begin{pmatrix}
 non\\ linear\\ term   
\end{pmatrix} 
\end{align}

In this case, using CMT, we can conclude that \( X \), \( Y \), \( Z \), and \( R \) are the stable variables, while \( L \) is the central variable. At this critical point, the matrices \( \mathcal{A} \) and \( \mathcal{B} \) appear as follows:

\[
\mathcal{A} =\left(
\begin{array}{ccccc}
-\frac{3 \sqrt{\left(\beta ^2 \eta ^2+6\right) (\beta  \eta  (8 \alpha +\beta  \eta )+6)}}{2 \left(\beta ^2 \eta ^2+6\right)}-\frac{3}{2} &0& 0& 0  \\
 0  &  \frac{3}{2} \Big[\frac{\sqrt{\left(\beta ^2 \eta ^2+6\right) (\beta  \eta  (8 \alpha +\beta  \eta )+6)}}{\beta ^2 \eta ^2+6}-1\Big]& 0 & 0 \\
0 & 0 & -3 & 0  \\
 0 & 0 & 0 & 0-2  \\
\end{array}
\right)
\hspace{0.5cm}
\mathcal{B} = 
\begin{bmatrix}
0    
\end{bmatrix}
\]

In this case the stable variables are \( X = g_{1}(L) \), \( Y = g_{2}(L) \), \( Z = g_{3}(L) \), and \( R = g_{4}(L) \), the zeroth approximation of the manifold functions as follows:

\begin{eqnarray}
\mathcal{N}(g_1(L)) = -\frac{3 \sqrt{6} L}{\beta ^2 \eta ^2+6}\,, \hspace{0.5cm} \mathcal{N}(g_2(L)) =\frac{3 \beta  \eta  L}{\beta ^2 \eta ^2+6}\,, \hspace{0.5cm} \mathcal{N}(g_3(L)) =0 \,,\hspace{0.5cm} \mathcal{N}(g_4(L)) =0 \,
\end{eqnarray}

With these assumptions, the central manifold can be obtained as follows:
 
\begin{equation}\label{CMTmanifold4}
\dot{L}=\frac{36 \beta  L^2}{\beta ^2 \eta ^2+6}+ higher \hspace{0.2cm} order \hspace{0.2cm} term   
\end{equation}

According to CMT, this critical point exhibits stable behavior for \( \eta \in \mathbb{R} \) and \( \beta < 0 \), where \( \dot{L} \) is negative.
\\

\textbf{CMT for critical point $\mathcal{G}_{DE}$:}\\

The Jacobian matrix at the critical point $\mathcal{G}_{DE}$ for the autonomous system (\ref{dynamicalsystem} ) is as follows:

$J(\mathcal{G}_{DE}) = 
\left(
\begin{array}{ccccc}
 -3 & 0 & -3 \sqrt{\frac{3}{2}} & 0 & \sqrt{\frac{3}{2}} \\
 0 & -3 & 0 & 0 & 0 \\
 0 & 0 & 0 & 0 & 0 \\
 0 & 0 & 0 & -2 & 0 \\
 \sqrt{6} \beta ^2 \eta  & 0 & 0 & 0 & 0 \\
\end{array}
\right)$

At the critical point $\mathcal{G}_{DE}$ we have the eigenvalues $\nu_{1}=-3,\,\nu_{2}=-2,\,\nu_{3}=0,\,\nu_{4}=\frac{1}{2} \left(-\sqrt{12 \beta ^2 \eta +9}-3\right),\,\nu_{5}=\frac{1}{2} \left(\sqrt{12 \beta ^2 \eta +9}-3\right)$ The eigenvectors corresponding to these eigenvalues are 
\begin{math}
   \left[0,1,0,0,0\right]^{T} 
\end{math},\\
\begin{math}
   \left[0,0,0,1,0\right]^{T} 
\end{math}
\begin{math}
   \left[0,0,\frac{1}{3},0,1\right]^{T} 
\end{math}
\begin{math}
   \left[-\frac{\sqrt{2} \sqrt{4 \beta ^2 \eta +3}+\sqrt{6}}{4 \beta ^2 \eta },0,0,0,1\right]^{T} 
\end{math}
\begin{math}
\left[-\frac{\sqrt{6}-\sqrt{2} \sqrt{4 \beta ^2 \eta +3}}{4 \beta ^2 \eta },0,0,0,1\right]^{T} 
\end{math}

In this case, to shift the critical point towards the origin, the transformations we used are: \( X = x \), \( Y =  y-1 \), \( Z = u \), \( R = \rho \), and \( L = \lambda \). The equations in the new co-ordinate system can be written as:

\begin{align}
\begin{pmatrix}
\dot{X}\\ 
\dot{Y} \\ 
 \dot{R}\\ 
 \dot{L} \\
 \dot{Z}
\end{pmatrix}= 
\left(
\begin{array}{ccccc}
\frac{1}{2} \left(-\sqrt{12 \beta ^2 \eta +9}-3\right)&0& 0& 0 &0 \\
 0  & \frac{1}{2} \left(\sqrt{12 \beta ^2 \eta +9}-3\right)& 0 & 0 &0\\
0 & 0 & -3 & 0 & 0 \\
 0 & 0 & 0 & 0-2 & 0 \\
 0 & 0 & 0 & 0 & 0 \\
\end{array}
\right)
\begin{pmatrix}
X\\ 
 Y\\ 
 R\\ 
 L\\ 
 Z
\end{pmatrix}+\begin{pmatrix}
 non\\ linear\\ term   
\end{pmatrix} 
\end{align}

In this case the stable variables are \( X \), \( Y \), \( R \), and \( L \) , while \( Z \) is the central variable. At this critical point, the matrices \( \mathcal{A} \) and \( \mathcal{B} \) appear as follows:

\[
\mathcal{A} =\left(
\begin{array}{ccccc}
\frac{1}{2} \left(-\sqrt{12 \beta ^2 \eta +9}-3\right) &0& 0& 0  \\
 0  &  \frac{1}{2} \left(\sqrt{12 \beta ^2 \eta +9}-3\right)& 0 & 0 \\
0 & 0 & -3 & 0  \\
 0 & 0 & 0 & 0-2  \\
\end{array}
\right)
\hspace{0.5cm}
\mathcal{B} = 
\begin{bmatrix}
0    
\end{bmatrix}
\]
 We have assumed the following functions for the stable variables: \( X = g_{1}(Z) \), \( Y = g_{2}(Z) \), \( R = g_{3}(Z) \), and \( L = g_{4}(Z) \). The zeroth approximation of the manifold functions can be obtained as follows:

\begin{eqnarray}
\mathcal{N}(g_1(Z)) =\frac{3 \sqrt{6} Z}{3 Z^2+2}\,, \hspace{0.5cm} \mathcal{N}(g_2(Z)) =-\frac{9 Z^2}{3 Z^2+2}\,, \hspace{0.5cm} \mathcal{N}(g_3(Z)) =0 \,,\hspace{0.5cm} \mathcal{N}(g_4(Z)) =0 \,
\end{eqnarray}

With these, the central manifold can be obtained as follows:
 
\begin{equation}\label{CMT-manifold}
\dot{Z}=-\frac{18 \alpha  Z^2}{3 Z^2+2}+ higher \hspace{0.2cm} order \hspace{0.2cm} term   
\end{equation}

According to CMT, this critical point exhibits stable behavior for \( Z \ne 0 \) and \( \alpha > 0 \), where \( \dot{Z} \) is negative.
\\

\section*{Acknowledgements}
SAK and LKD acknowledges the financial support provided by the University Grants Commission (UGC) through Senior Research Fellowship (UGC Ref. No.: 191620205335), and (UGC Ref. No.: 191620180688) respectively. BM acknowledges SERB-DST to provide financial support under the MATRICS grant (MTR/2023/000371). 
\section*{References} 
\bibliographystyle{utphys}
\bibliography{references}

\end{document}